\documentclass[]{pasj02} 
\usepackage[switch,mathlines]{lineno} 
\usepackage{natbib} 

\jyear{2024}
\Received{2024/09/18}
\Accepted{2024/11/20}

\newcommand{\beq}{\begin{equation}}
\newcommand{\eeq}{\end{equation}}
\newcommand{\beqn}{\begin{eqnarray}}
\newcommand{\eeqn}{\end{eqnarray}}
\newcommand{\pfrac}[2]{ \biggl(\dfrac{#1}{#2}\biggr) }

\newcommand{\pd}{\partial}
 
\newcommand{\kB}{k_{\rm B}}

 \newcommand{\brac}[1]{\langle #1 \rangle}

\newcommand{\sun}{\odot}

\begin{document}

\title{Surface accretion as a dust retention mechanism in protoplanetary disks. I.  Formulation and proof-of-concept simulations}

\author{
Satoshi \textsc{Okuzumi}\altaffilmark{1}\altemailmark\orcid{0000-0002-1886-0880}\email{okuzumi@eps.sci.titech.ac.jp}
}

\altaffiltext{1}{Department of Earth and Planetary Sciences, Institute of Science Tokyo, Meguro, Tokyo 152-8551, Japan}



\maketitle

\begin{abstract}
Planetesimal formation via the streaming and gravitational instabilities of dust in protoplanetary disks requires a local enhancement of the dust-to-gas mass ratio. Radial drift of large grains toward pressure bumps in gas disks is a plausible mechanism for achieving the required dust concentration. However, recent millimeter disk observations suggest that the maximum sizes of dust grains in these disks are considerably smaller than predicted by dust evolution models that assume sticky grains. This indicates that the grains may be more strongly coupled to the gas and hence drift more slowly than previously anticipated. In this study, we propose a new dust retention mechanism that enables an enhancement of the dust-to-gas mass ratio in disks with slowly drifting grains. This mechanism assumes that a surface accretion flow driven by magnetohydrodynamical (MHD) winds removes disk gas while retaining the slowly drifting grains below the flow. This process is expected to occur when the timescale of gas removal is shorter than the timescale of dust radial advection. To test this, we develop a radially one-dimensional framework for the transport of gas and dust in a disk with a vertically nonuniform accretion structure. Using this framework, we simulate the growth, fragmentation, and radial transport of dust grains in surface-accreting disks. Our simulations confirm a significant enhancement of the midplane dust-to-gas mass ratio when the predicted conditions for dust retention are met. Dust retention by MHD-driven surface accretion flows may thus pave the way for planetesimal formation from poorly sticky grains.

\end{abstract}


\section{Introduction}\label{sec:intro}
How dust grains in protoplanetary disks form kilometer-sized planetesimals remains an outstanding question in planet formation theory. It is widely accepted that (sub)micron-sized grains in these disks coagulate through mutual sticking, forming macroscopic aggregates, often called ``pebbles'' \citep{Johansen14}. However, the extent to which these aggregates can grow is much less clear, as their collision velocities often exceed 1--10 $\rm m~s^{-1}$, potentially leading to fragmentation or bouncing rather than sticking upon collision \citep[e.g.,][]{Guttler10,Zsom11,BirnstielKlahr12,DominikDullemond24}. In addition, aerodynamic drag causes macroscopic aggregates to drift radially, either inward or outward, relative to the background gas \citep{Whipple72,Adachi76,Weidenschilling77a}. Growth of the aggregates can only occur if they grow faster than they drift. If the grains are extremely sticky, they may form highly porous aggregates and grow faster than they drift \citep{Okuzumi12,Kataoka13b}. Otherwise, the aggregates can only grow to the maximum size limited by bouncing, fragmentation, or radial drift \citep{BirnstielKlahr12}. In this case, planetesimal formation is expected to occur via the gravitational collapse of dust overdensities produced by dust settling and radial pile-up \citep[e.g.,][]{GoldreichWard73,Sekiya98,YoudinShu02,IdaGuillot16,HyodoIdaGuillot21}, the streaming instability \citep[e.g.,][]{YoudinGoodman05,JohansenYoudin07,Carrera15,Yang17,LiYoudin21}, pressure bumps \citep[e.g.,][]{Whipple72,HaghighipourBoss03,Pinilla12}, vortices \citep[e.g.,][]{BargeSommeria95}, or secular gravitational instability \citep[e.g.,][]{Youdin11,TakahashiInutsuka14,Tominaga20}. The stickiness of dust grains dictates which pathway of planetesimal formation is likely to be realized. 

However, the stickiness of real dust grains in protoplanetary disks is poorly understood from both theoretical and experimental perspectives. In principle, the stickiness of grains depends on their size, surface roughness, and material composition \citep{Dominik97}. Until recently, it was believed that grains coated with water ice are so sticky that they can stick at collision velocities up to 10--70 $\rm m~s^{-1}$, depending on the size of the grains constituting the aggregates \citep{Dominik97,Wada09,Wada13,Gundlach15}. This led to the conventional idea that grains in the outer regions of protoplanetary disks, which are presumably ice-rich, could grow to sizes ranging from centimeters to even decimeters \citep[e.g.,][]{Birnstiel10a,Okuzumi12}. However, more recent experiments \citep{Gundlach18,Musiolik19} suggest that water ice is not as sticky at low temperatures ($\lesssim$150--200 K) as previously thought. Additionally, other experiments \citep{Musiolik16a,Musiolik16b,Fritscher21} indicate that CO$_2$ ice, which likely exists in disk regions with temperatures below 80 K \citep[e.g.,][]{Okuzumi16}, is less sticky than water ice (see \citealt{ArakawaKrijt21} for a possible explanation of why CO$_2$ ice is less sticky). Clearly, further theoretical and experimental investigations are needed to fully understand the sticking properties of dust grains of various compositions.

On the other hand, growing observational evidence suggests that grains in the outer parts of protoplanetary disks are indeed less sticky than previously expected. Multiwavelength and polarimetric observations of dust thermal emission at (sub)millimeter wavelengths provide information about the size distribution of dust grains/aggregates in disks \cite[for a review, see][]{Miotello23}. Uniformly polarized (sub)millimeter emission from a number of disks \cite[e.g.,][]{Stephens17,Stephens23,Hull18} is widely interpreted as evidence for $\sim0.1$--1 mm-sized grains being abundant, at least in the outer parts of these disks \citep{Kataoka15,Yang16,UedaKataoka21}. The necessity for abundant 0.1--1 mm-sized grains also aligns with the spectral slopes of the (sub)millimeter emission \citep{Liu19,Zhu19,Ueda20,Chung24}. Since turbulence in the outer regions of protoplanetary disks is generally weak \citep[for a review, see][]{Rosotti23}, the dominance of 0.1--1 mm-sized grains suggests that they are poorly sticky; otherwise, they would grow larger in such a quiescent environment. These observations are consistent with dust evolution models that assume poorly sticky grains, with maximum sticking velocities as low as 0.2--1 $\rm m~s^{-1}$ \citep{Okuzumi19,Jiang24,Ueda24}.

The possibility that dust grains in protoplanetary disks may be poorly sticky has significant implications for our understanding of dust evolution and planetesimal formation. Firstly, it seems extremely unlikely that dust coagulation alone leads to planetesimal formation. Secondly, since poorly sticky grains stop growing while still being strongly coupled to the gas disk, they would only drift radially at a slow speed. 
Interestingly, millimeter disk observations show that the dust disks in the $\sim$5-Myr-old Upper Scorpius star-forming region are, on average, smaller than those in $\sim$1--2 Myr-old star-forming regions, but only by a factor of $\sim$2 \citep{Hendler20}. This could potentially indicate that the grains only drift on a timescale of several Myr. In comparison, conventional sticky icy grains would grow to 1--10 cm in size and drift toward the central star within 1 Myr \citep[e.g.,][]{BirnstielKlahr12,Okuzumi12}, unless they are trapped by pressure bumps \citep{Pinilla12}. Recently, \citet{Lee24} analyzed the masses of dust rings found in recent millimeter disk observations and concluded that the grains would be considerably coupled to the gas, meaning they would drift inward only slowly. This is because the inferred dust ring masses are considerably smaller than what would be expected if the pressure bumps had already trapped all the grains in the outer disks.

A low speed of dust radial drift could severely limit the potential pathways for planetesimal formation. Both runaway dust settling toward the midplane \citep{Sekiya98,YoudinShu02} and strong dust clumping by the streaming instability \citep{Carrera15,Yang17,LiYoudin21} require a prior enhancement of the dust-to-gas mass ratio. Radial dust drift toward pressure bumps has been considered a plausible mechanism for local dust concentration leading to planetesimal formation \citep[e.g.,][]{HaghighipourBoss03,Johansen14}. However, slow radial drift delays this concentration process. The analysis by \citet{Lee24} suggests that the observed dust rings are stable against strong clumping by the streaming instability.

Even if radial dust concentration is inefficient, increasing the disk's dust-to-gas mass ratio is possible by removing gas. Previous studies have proposed that gas removal by photoevaporative \citep{Gorti15,Carrera17} and magnetohydrodynamical (MHD) disk winds \citep{Suzuki10,Bai16} can potentially lead to planetesimal formation. However, the efficiency of planetesimal formation via wind mass loss critically depends on the assumed mass loss rate and its radial profile \citep{Ercolano17}.

In this study, we propose surface gas accretion driven by MHD winds as an alternative mechanism for dust retention in disks with slowly drifting grains. Global non-ideal MHD simulations of protoplanetary disks show that gas accretion is narrowly concentrated on the surface of inner disk regions having a poorly ionized interior \citep[e.g.,][]{Bai13,Bai17,BaiStone13b,Gressel15,Riols20,Lesur21,Iwasaki24}. The emergence of this surface accretion flow is a consequence of the disk surface being relatively well-ionized and, therefore, well-coupled to the magnetic field threading the disk. The wind, driven by the global magnetic field, extracts angular momentum from the disk surface, driving a surface accretion flow. This flow transports gas, but will not transport dust that has already settled on the midplane due to stellar gravity. In this study, we demonstrate that surface gas accretion, combined with slow radial dust drift, can indeed lead to the enhancement of the dust-to-gas mass ratio required for planetesimal formation.

\begin{figure*}[t]
\includegraphics[width=\hsize, bb=0 0 1920 980]{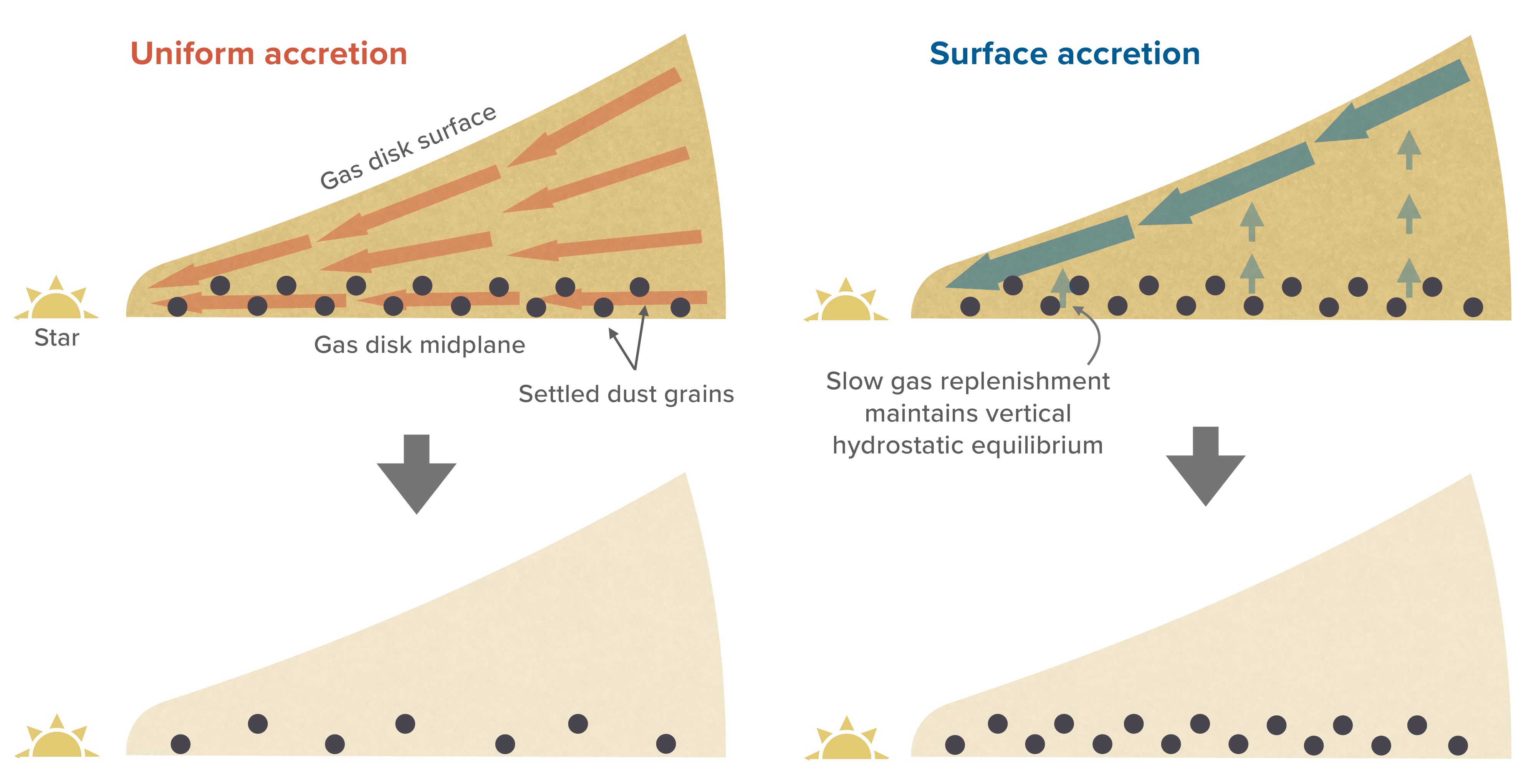}
\caption{Schematic showing gas and dust transport in protoplanetary disks with uniform and surface accretion flows (left and right figures, respectively). In a disk with a vertically uniform accretion flow, the gas flow around the midplane flushes the vertically settled dust toward the central star. In a disk with an accretion flow localized near its surface, the settled dust avoids accretion toward the star. As long as the disk gas remains approximately in vertical hydrostatic equilibrium, the decrease in gas surface density due to surface accretion always results in a decrease in midplane gas density (see text). Therefore, the dust-to-gas ratio at the midplane in the surface-accreting disk would increase over time, provided that the radial inward drift of the dust due to gas drag is slow.
}
\label{fig:flow}
\end{figure*}

This paper is organized as follows. Section~\ref{sec:idea} presents the basic mechanism of dust retention proposed in this study. We then formulate the radial transport of gas and dust in surface-accreting disks in section~\ref{sec:formulation}. Our  proof-of-concept simulation models are presented in section~\ref{sec:model} and their results are provided in section~\ref{sec:results}. Our conclusions are presented in section~\ref{sec:summary}. 

\section{Dust retention in surface-accreting disks: the mechanism}~\label{sec:idea}
We begin with thought experiments on the radial transport of dust grains in an accretion disk (figure~\ref{fig:flow}). {We assume that the grains have already grown and settled to the midplane \citep[e.g.,][]{Weidenschilling80,Nakagawa81}.} These grains do not only accrete toward the central star with the midplane gas but also drift relative to the gas due to gas drag \citep{Whipple72,Adachi76,Weidenschilling77a,Takeuchi02}. If there are no pressure bumps, the drift velocity is negative, and therefore the grains' net inward velocity is higher than that of the gas at the midplane. 
The question is whether the dust-to-gas surface density ratio will then decrease or increase. As illustrated by the two examples described below, the answer depends on the vertical distribution of the accretion flow. To isolate the role of accretion flows, we ignore any gas removal by disk winds in this section.

First, we consider the simplest case of a vertically uniform accretion flow (the left side of figure~\ref{fig:flow}). 
In this case,  both the gas and dust mass accretion fluxes are dominated by the flow at the midplane, where their densities are the highest.
Because the dust accretes faster than the gas, the disk's dust-to-gas mass ratio will decrease over time.

Next, we consider a disk with a strong accretion flow near its surface (the right side of figure~\ref{fig:flow}), as is often observed in non-ideal MHD simulations of wind-driven accretion. In this case, dust accretion at the midplane is independent of the surface gas accretion, allowing the dust accretion flux to be smaller than the gas accretion flux. When this occurs, the disk's dust-to-gas surface density ratio will increase over time. 

One might ask whether surface accretion will also lead to an increase in the dust-to-gas density ratio {\it at the midplane}, which is likely more relevant to planetesimal formation than the dust-to-gas surface density, as the streaming and gravitational instabilities of dust overdensities occur at the midplane \citep[e.g.,][]{Gole20,LiYoudin21}. We expect that a decrease in the gas surface density due to surface accretion will also lead to a decrease in the midplane gas density, provided that gas depletion is slow enough for the disk to maintain vertical hydrostatic equilibrium. When a surface accretion flow removes gas near the disk surface, the decreased surface gas pressure yields an excess upward pressure gradient force. This excess force drives an upward gas flow, transporting gas from lower to higher altitudes to restore vertical hydrostatic equilibrium (see figure~\ref{fig:flow}). {Consequently, the midplane gas density will decrease with gas surface density}\footnote{Equation~\eqref{eq:rhog_Gaussian} in the main text explicitly shows that, under vertical hydrostatic equilibrium, the midplane gas density is proportional to the gas surface density.}. {It is important to note here that this upward replenishing flow is slow, moving from the midplane to the surface over the timescale on which the gas surface density decreases---namely, the gas disk's lifetime. This flow will have little impact on the vertical distribution of the already settled dust grains, as their settling timescale is shorter than the disk's lifetime. This means that the upward flow will change the midplane gas density but {\it not} the midplane dust density.}

Therefore, we expect that the midplane dust-to-gas density ratio in a surface-accreting disk will increase if the dust at the midplane drifts more slowly than the gas surface density decreases. Once the midplane dust-to-gas density ratio exceeds unity, runaway dust settling  \citep[e.g.,][]{Sekiya98,YoudinShu02}, runaway pile-up of drifting dust \citep[e.g.][]{Drazkowska16,IdaGuillot16,HyodoIdaGuillot21}, or strong dust clumping via the streaming instability \citep[e.g.,][]{YoudinGoodman05,Sekiya18,Gole20,LiYoudin21} will lead to the gravitational collapse of the dust overdensities into planetesimals. Since the drift speed of small grains increases with their size, the dust retention mechanism requires the grains at the midplane to be sufficiently small (but still large enough for significant settling to occur). This requirement can be fulfilled if the grains are poorly sticky.

The surface-accreting disk considered here is motivated by recent non-ideal MHD simulations of protoplanetary disks. These simulations typically show strong accretion flows near the base of MHD winds, lying at 2--4 scale heights above and/or below the midplane, in a inner disk region with a high magnetic resistivity (e.g., figure 7 of \citealt{Gressel15}; figures 4 and 9 of \citealt{Bai17}; figure 13 of \citealt{Lesur21}; figure 34 of \citealt{Iwasaki24}). The observed surface flows comprise a significant fraction of the vertically integrated mass accretion rate. We note that strong surface accretion flows do not necessarily emerge everywhere in the disks; in outer disk regions with a lower magnetic resistivity, strong accretion tends to occur closer to the midplane \citep{Bethune17,Bai17,Suriano18,Suriano19,Lesur21,CuiBai21,Iwasaki24}. Lacking a generic model for the radial vertical structure of MHD-driven gas accretion flows, we choose to model the radial transport of gas and dust in accretion disks with vertically nonuniform accretion structures using vertically integrated transport equations, which are formulated in section~\ref{sec:formulation}.

We also note that some MHD simulations show a pair of accretion and decretion flows around the midplane \citep{Bai17,CuiBai21}. These flows are nearly anti-symmetric and largely cancel each other out with respect to the net accretion rate. However, they can drive strong radial diffusion of vertically well-mixed dust \citep{HuBai21}. We discuss the potential effect of such strong radial mixing on dust retention in section~\ref{sec:Dr}.

\section{Formulating gas and dust transport in disks with vertically nonuniform accretion structures} \label{sec:formulation}
The goal of this work is to demonstrate dust retention in surface-accreting disks. To this end, we need to treat the radial transport of gas and dust while accounting for the vertical distribution of their densities and accretion velocities. In this section, we formulate this problem by vertically integrating the  equations of continuity for gas and dust.

In the following, we consider an axisymmetric disk and use the standard cylindrical coordinate system $(r, \phi, z)$, with $z = 0$ corresponding to the disk's midplane.
\subsection{Gas transport}

The equation of continuity for gas in an axisymmetric disk can be expressed as
\begin{equation}
\frac{\pd\rho_{\rm g}}{\pd t} 
= -\frac{1}{r}\frac{\pd}{\pd r}
(r {v_{{\rm g},r}}\rho_{\rm g} )
 - \frac{\pd }{\pd z} (v_{{\rm g},z}\rho_{\rm g}),
 \label{eq:rhog}
\end{equation}
where $\rho_{\rm g}(r,z)$, $v_{{\rm g},r}(r,z)$, and $v_{{\rm g},z}(r,z)$ are the density, radial velocity, and vertical velocity of the gas, respectively. Integrating equation~\eqref{eq:rhog} over the full vertical extent of the disk yields
\beq
\frac{\pd\Sigma_{\rm g}}{\pd t} 
= -\frac{1}{r}\frac{\pd}{\pd r}
[r \brac{v_{{\rm g},r}}_{\rm g}\Sigma_{\rm g} ]
- \dot{\Sigma}_{\rm g}^{\rm wind}.
\label{eq:Sigmag}
\eeq
where $\Sigma_{\rm g}(r) = \int \rho_{\rm g}(r,z) dz$ is the gas surface density, $\dot{\Sigma}_{\rm g}^{\rm wind}$ is the rate of gas surface density loss from the disk surface per unit time, and $\brac{v_{{\rm g},r}}_{\rm g}(r)$ is the gas-density-weighted average of $v_{{\rm g},r}(r,z)$. For any field $X(r,z)$, its gas-density-weighted average is defined as 
\beq
\brac{X}_{\rm g}(r) 
\equiv \frac{1}{\Sigma_{\rm g}(r)}\int X(r,z) \rho_{\rm g}(r,z) dz
\label{eq:ovlX}
\eeq

A key point to note here is that the mass-weighted average velocity $\brac{v_{{\rm g},r}}_{\rm g}$ governs the advective transport of $\Sigma_{\rm g}$. Depending on the vertical profile of $v_{{\rm g},r}$, the  value of $\brac{v_{{\rm g},r}}_{\rm g}$ is not necessarily close to the velocity at the midplane, $v_{{\rm g},r}(z=0)$.
For instance, when the accretion flow is localized near the disk surface, as shown on the right side of figure 1,  $|\brac{v_{{\rm g},r}}_{\rm g}|$ would be significantly smaller than $|v_{{\rm g},r}(z)|$ at the accretion surface and significantly larger than $|v_{{\rm g},r}(z=0)|$.

An order-of-magnitude estimate of equation~\eqref{eq:Sigmag} shows that gas advective transport occurs on a timescale of  
\beq
t_{\rm adv,g} \approx \frac{r}{|\langle v_{{\rm g},r}\rangle_{\rm g}|}.
\label{eq:tadvg}
\eeq
\subsection{Dust transport}\label{sec:transport_dust}
For the dust component, we account for both advection and turbulent diffusion within the gas disk. The equation of continuity for the dust is given by
\begin{align}
\frac{\pd\rho_{\rm d}}{\pd t} =& -\frac{1}{r}\frac{\pd}{\pd r}
r\left[{v_{{\rm d},r}}\rho_{\rm d} -\rho_{\rm g} D_{{\rm d},r}\frac{\pd}{\pd r}\pfrac{\rho_{\rm d}}{\rho_{\rm g}} \right] 
\nonumber \\
&- \frac{\pd }{\pd z} \left[v_{{\rm g},z}\rho_{\rm d}-\rho_{\rm g}D_{{\rm d},z}\frac{\pd}{\pd z}\pfrac{\rho_{\rm d}}{\rho_{\rm g}}  \right],
 \label{eq:rhod}
\end{align}
where $\rho_{\rm d}(r,z)$ is the dust density, $v_{{\rm d},r}(r,z)$ and $v_{{\rm d},z}(r,z)$ are the radial and vertical dust velocities, and $D_{{\rm d},r}(r,z)$ and $D_{{\rm d},z}(r,z)$ are the radial and vertical dust diffusion coefficients, respectively. Vertical integration of equation~\eqref{eq:rhod} yields
\begin{align}
\frac{\pd\Sigma_{\rm d}}{\pd t} 
=& -\frac{1}{r}\frac{\pd}{\pd r}r\left[
\brac{ v_{{\rm d},r}}_{\rm d}\Sigma_{\rm d}
- \Sigma_{\rm g}\left\langle D_{{\rm d},r}\frac{\pd}{\pd r}\pfrac{\rho_{\rm d}}{\rho_{\rm g}} \right\rangle_{\rm g}
\right]
\notag \\
& - \dot{\Sigma}_{\rm d}^{\rm wind},
\label{eq:Sigmad0}
\end{align}
where $\Sigma_{\rm d}(r)$ is the dust surface density, $ \dot{\Sigma}_{\rm d}^{\rm wind}$ is the rate of dust surface density less from the disk surface per unit time, and $\brac{v_{{\rm d},r}}_{\rm d}$ is the dust-density-weighted vertical average of $v_{{\rm d},r}$. The dust-density-weighted average of any field $X(r,z)$ is generally defined as  
\beq
\brac{X}_{\rm d}(r)
\equiv \frac{1}{\Sigma_{\rm d}(r)}\int X(r,z) \rho_{\rm d}(r,z) dz.
\eeq 
Note that $\brac{X}_{\rm g} \not= \brac{X}_{\rm d}$ when the dust is settled around the midplane. 
The averaging $\langle D_{{\rm d},r}{\pd(\rho_{\rm d}/\rho_{\rm g})}/{\pd r} \rangle_{\rm g}$ left in equation~\eqref{eq:Sigmad0} can be performed if the vertical profiles of $\rho_{\rm g}$, $\rho_{\rm d}$, and $D_{{\rm d},r}$ are specified. We defer this task to section~\ref{sec:grains} and appendix~\ref{sec:diff}. 

If the dust consists of grains of different sizes, the radial dust velocity should also be averaged over the grain size distribution. For the sake of clarity, the following formulation assumes equally sized grains.

The radial dust velocity is induced by co-accretion with the gas \citep{Takeuchi02} and radial drift due to gas drag \citep{Whipple72,Adachi76,Weidenschilling77a}. This velocity depends on the dust grains' stopping time $t_{\rm stop}$, or equivalently their Stokes number ${\rm St} \equiv  \Omega t_{\rm stop}$, where $\Omega$ is the local Keplerian frequency. Neglecting aerodynamical feedback from dust to gas \citep{Nakagawa86,Kretke09,DipierroLaibe17,KanagawaUeda17}, the gas radial velocity can be expressed as \citep{Takeuchi02}
\beq
{ v_{{\rm d},r}} = 
\frac{  v_{{\rm g},r} }{1+{\rm St}^2} + \frac{2 {\rm St}\, \Delta v_{{\rm g},\phi}}{1+{\rm St}^2},
\label{eq:vdravr}
\eeq
where $\Delta v_{{\rm g},\phi}$ is the gas rotation velocity relative to Keplerian ($\Delta v_{{\rm g},\phi} <0$ for sub-Keplerian motion). On the right-hand side of equation~\eqref{eq:vdravr}, the first and second terms represent the contributions from co-accretion and drift, respectively. Since we neglect the feedback of dust on gas, the vertical variation of $\Delta v_{{\rm g},\phi}$ within respect to $z$ is insignificant \citep{Takeuchi02}.
Therefore, the vertical average of the second term in equation~\eqref{eq:vdravr} can be approximated by its midplane value, allowing us to express $\brac{ v_{{\rm d},r}}_{\rm d}$ as
\beq
\brac{ v_{{\rm d},r}}_{\rm d}
\approx
\left\langle\frac{v_{{\rm g},r}}{1+{\rm St}^2} \right\rangle_{\rm d}
+ \frac{2 {\rm St}_{\rm mid}\Delta v_{{\rm g},\phi,{\rm mid}}}{1+{\rm St}_{\rm mid}^2},
\label{eq:vdravr_approx}
\eeq
where the subscript ``mid'' denotes a midplane value. 

By analogy with equation~\eqref{eq:tadvg}, the timescale of dust advective transport is given by 
\beq
t_{\rm adv,d} \approx \frac{r}{|\langle v_{{\rm d},r}\rangle_{\rm d}|}.
\label{eq:tadvd}
\eeq
If $\langle v_{{\rm g},r}\rangle_{\rm g}$ and $\langle v_{{\rm d},r}\rangle_{\rm d}$ are negative, then $t_{\rm adv,g}$ and $t_{\rm adv,d}$ represent the timescale of gas and dust removal from the disk, respectively. It follows that dust will be lost more slowly than gas (i.e., $t_{\rm adv,d} > t_{\rm adv,g}$) if 
\beq
|\langle v_{{\rm d},r}\rangle_{\rm d}| < |\langle v_{{\rm g},r}\rangle_{\rm g}|.
\label{eq:cond_dustretention}
\eeq
Dust retention is expected to occur when equation \eqref{eq:cond_dustretention} is satisfied.

\subsection{Impact of vertical accretion structure on dust transport: examples}
\label{sec:examples}

The first term $\langle {v_{{\rm g},r}}/({1+{\rm St}^2}) \rangle_{\rm d}$ on the right-hand side of equation~\eqref{eq:vdravr_approx} encapsulates the effect of the vertical distribution of $v_{{\rm g},r}(z)$ on radial dust advection.
We illustrate this effect using the two extreme examples shown in figure~\ref{fig:flow}.

\subsubsection{Uniform accretion}

If gas accretion is vertically uniform, as illustrated on the left side of figure~\ref{fig:flow}, we have $v_{{\rm g},r}(z) = \brac{v_{{\rm g},r}}_{\rm g}$ for all $z$, and hence 
\beq
\left\langle\frac{v_{{\rm g},r}}{1+{\rm St}^2} \right\rangle_{\rm d} 
\approx \frac{\brac{v_{{\rm g},r}}_{\rm g}}{1+{\rm St}^2_{\rm mid}},
\label{eq:coacc_uniform}
\eeq
where have used $\langle 1/({1+{\rm St}^2}) \rangle_{\rm d} \approx 1/({1+{\rm St}_{\rm mid}^2})$. 
This is the standard expression for the dust co-accretion velocity used in the literature. 

We are particularly interested in 0.1--1 mm-sized grains, which generally fulfill ${\rm St}_{\rm mid} \ll 1$. In this case, equation~\eqref{eq:vdravr_approx} approximates to
\beq
\brac{ v_{{\rm d},r}}_{\rm d} \approx 
 \brac{v_{{\rm g},r}}_{\rm g}
+ {2 {\rm St}_{\rm mid}\Delta v_{{\rm g},\phi,{\rm mid}}}.
\label{eq:vdravr_uniform_approx}
\eeq
For accretion disks with sub-Keplerian rotation, both the first and second terms on the right-hand side of equation \eqref{eq:vdravr_uniform_approx} are negative. Therefore, equation~\eqref{eq:cond_dustretention} never holds, meaning that dust will be lost from the disk faster than gas.

\subsubsection{Surface accretion}

We now consider surface accretion flow like that depicted on the right side of figure~\ref{fig:flow}. To make vertical averaging analytically tractable, we assume  this flow to be localized on a thin layer at a height of $z=z_{\rm s}$ above the midplane. As shown in appendix~\ref{sec:local}, the vertically averaged co-accretion term $\langle {v_{{\rm g},r}}/({1+{\rm St}^2}) \rangle_{\rm d}$ for this case can be written as 
\beq
\left\langle\frac{v_{{\rm g},r}}{1+{\rm St}^2} \right\rangle_{\rm d} 
\approx C_{\rm surface}
\frac{ \brac{v_{{\rm g},r}}_{\rm g} }{1+{\rm St}(z_s)^2},
\label{eq:coacc_surface}
\eeq
where 
\beq
C_{\rm surface} \equiv \frac{\rho_{\rm d}(z_{\rm s})/\rho_{\rm g}(z_{\rm s})}{\Sigma_{\rm d}/\Sigma_{\rm g}}
\label{eq:Csurface_def}
\eeq
is a normalized dust-to-gas ratio at the accretion layer. This coefficient equals unity if dust is well-mixed up to the layer, but vanishes if dust is heavily depleted from the layer. In appendix~\ref{sec:local} we derive an analytic expression for $C_{\rm surface}$ for the special case where the vertical dust  distribution is determined by the balance between settling and (vertically uniform) diffusion. 
The result, presented in  equation~\eqref{eq:Csurface} and figure~\ref{fig:Csurface}, shows that $C_{\rm surface}$ indeed vanishes as the grains settle below the accretion surface.

If ${\rm St}_{\rm mid}(z_{\rm s}) < 1$, equation~\eqref{eq:vdravr_approx} for the surface accretion disk approximates to
\beq
\brac{ v_{{\rm d},r}}_{\rm d} \approx 
  C_{\rm surface}\brac{v_{{\rm g},r}}_{\rm g}
+ {2 {\rm St}_{\rm mid}\Delta v_{{\rm g},\phi,{\rm mid}}}.
\label{eq:vdravr_surface_approx}
\eeq
In the ideal case of $C_{\rm surface} \approx 0$, equation~\eqref{eq:cond_dustretention} holds if 
\beq
{\rm St}_{\rm mid} < \frac{\langle v_{{\rm g},r}\rangle_{\rm g}}{2\Delta v_{{\rm g},\phi,{\rm mid}}}.
\label{eq:St_dustretention}
\eeq
Again, we have assumed that both $\langle v_{{\rm g},r}\rangle_{\rm g}$ and $\Delta v_{{\rm g},\phi,{\rm mid}}$ are negative.
Since the grains' Stokes number increases with their size, equation~\eqref{eq:St_dustretention} requires that the grains be sufficiently small, as expected in section~\ref{sec:idea}.

\section{Simulation models}\label{sec:model}
So far, we have predicted that surface accretion can lead to dust retention. In the remainder of this work, we test this prediction by performing radially one-dimensional simulations of gas and dust evolution. In this section, we describe our gas and dust evolution models.

The aim of our simulations is to examine whether surface accretion alone can lead to an enhancement of the dust-to-gas ratio above unity. For this reason, we continue to neglect the aerodynamic feedback from dust to gas. If included, this effect would slow down inward dust drift in regions where the dust-to-gas ratio exceeds unity, further promoting planetesimal formation \citep[e.g.,][]{Drazkowska16,IdaGuillot16,KanagawaUeda17}.   For the same reason, we do not account for the conversion of dust into planetesimals in regions with high dust-to-gas ratios.

\subsection{Gas disk structure}
We assume stellar radiation to be the dominant source of disk heating. In poorly ionized protoplanetary disks, MHD accretion heating is indeed negligible beyond a few au \citep{Mori21,Kondo23}. In a passively irradiated disk, the temperature $T$ well below the stellar irradiation surface is approximately vertically uniform \citep[e.g.,][]{Calvet91,Chiang97}. The balance between stellar irradiation and radiative cooling yields the disk interior temperature given by \citep[see, e.g.,][]{Chiang97,Okuzumi22}
\beq
T = \pfrac{f_\downarrow \sin(\theta_*)L_*}{4\pi r^2 \sigma_{\rm SB}}^{1/4}
\eeq
where $L_*$ is the stellar luminosity, $\theta_*$ is the grazing angle between the starlight and the irradiation surface, $f_\downarrow$ is the fraction of the starlight flux reprocessed downward, and $\sigma_{\rm SB}$ is the Stefan--Boltzmann constant.
In this study, we adopt $f_\downarrow = 0.5$, $L_* = 1L_\odot$, and $\sin(\theta_*) = 0.03$, yielding
$T \approx 140(r/\rm 1~au)^{-1/2}~\rm K$. The choice of $f_\downarrow = 0.5$ assumes no grain scattering \citep{Chiang97,Okuzumi22}. Our value for $\sin(\theta_*)$ is taken from a consistent calculation of disk temperature and irradiation surface structures by \citet{Okuzumi22}.
Assuming a water ice sublimation temperature of 160 K, our adopted temperature profile places the snow line at $r \approx 0.7 ~\rm au$. For simplicity, we ignore ice sublimation at the snow line and treat all grains in the computational domain as icy. 

Assuming vertical hydrostatic equilibrium, the vertical gas density profile is given by 
\beq
\rho_{\rm g}(z) = \frac{\Sigma_{\rm g}}{\sqrt{2\pi} H_{\rm g}}\exp\left(-\frac{z^2}{2H_{\rm g}^2}\right),
\label{eq:rhog_Gaussian}
\eeq
where $H_{\rm g} = c_{\rm s}/\Omega$ is the gas scale height, $c_{\rm s} = \sqrt{\kB T/m_{\rm g}}$ is the isothermal sound speed, $\kB$ is the Boltzmann constant, and $m_{\rm g}$ is the mean gas molecular mass. We assume a constant mean molecular weight of 2.3.
The Keplerian frequency is given by $\Omega = \sqrt{GM_*/r^3}$, where $G$ is the gravitational constant and $M_*$ is the stellar mass. 

From the radial force balance, the deviation of the gas rotation velocity from Keplerian velocity is given by \citep{Whipple72,Adachi76,Weidenschilling77a}
\beq
\Delta v_{{\rm g},\phi} =  \frac{c_{\rm s}^2}{2r\Omega}\frac{\pd\ln P}{\pd\ln r},
\label{eq:dvgphi}
\eeq
where $P = \rho_{\rm g}c_{\rm s}^2$ is the gas pressure.

\subsection{Viscosity, winds, and turbulence}
Disk accretion is assumed to be driven by both radial angular momentum transport within the disk and angular momentum removal by MHD winds. Physically, radial angular momentum transport may be caused by turbulence and/or horizontal magnetic fields amplified in the disk. In this study, we do not specify the physical origin of radial angular momentum transport and treat it using macroscopic viscosity, as in the standard viscous accretion disk model \citep{Lynden-Bell74}. Our accretion model essentially follows that of \citet{Tabone22a} but makes additional assumptions about the vertical structure of gas accretion flows (see section~\ref{sec:vertical}).

We decompose $v_{{\rm g},r}$ as
\begin{equation}
v_{{\rm g},r} = v_{{\rm g},r}^{\rm visc} + v_{{\rm g},r}^{\rm wind},
\end{equation}
where $v_{{\rm g},r}^{\rm visc}$ and $v_{{\rm g},r}^{\rm wind}$ denote radial gas velocities induced by the viscosity and MHD winds, respectively. 
Following \citet{Tabone22a}, we parameterize their gas-density-weighted averages as follows.
For $\brac{v_{{\rm g},r}^{\rm visc}}_{\rm g}$, we employ the expression from the viscous accretion model,
\beq
\brac{v_{{\rm g},r}^{\rm visc}}_{\rm g}
= - \frac{3\alpha_{\rm visc}c_{\rm s}^2}{r\Omega}\frac{\pd\ln(r^2 \alpha_{\rm visc}c_{\rm s}^2\Sigma_{\rm g})}{\pd \ln r},
\eeq
where $\alpha_{\rm visc}$ is the viscosity normalized by $c_{\rm s}^2/\Omega$ \citep{Shakura73}.
For $\brac{v_{{\rm g},r}^{\rm wind}}_{\rm g}$, we use 
\beq
\brac{v_{{\rm g},r}^{\rm wind}}_{\rm g}
= -\frac{3\alpha_{\rm wind}c_{\rm s}^2}{2r\Omega},
\label{eq:vgr_wind}
\eeq
where $\alpha_{\rm wind}$ is a dimensionless quantity characterizing the magnitude of the MHD wind stress, which removes the disk's angular momentum. The parameters $\alpha_{\rm visc}$ and $\alpha_{\rm wind}$ are equivalent to $\alpha_{\rm SS}$ and $\alpha_{\rm DW}$ of \citet{Tabone22a}, respectively.

MHD winds also cause mass loss from the disk. From angular momentum conservation, the gas mass loss rate $\dot{\Sigma}_{\rm g}^{\rm wind}$ in equation~\eqref{eq:Sigmag} can be written as 
\beq
\dot{\Sigma}_{\rm g}^{\rm wind} 
= \frac{3\alpha_{\rm wind}\Sigma_{\rm g}c_{\rm s}^2}{4(\lambda-1)\Omega r^2},
\eeq
where the dimensionless number $\lambda$, often called the magnetic lever arm parameter, is the ratio of the the total specific angular momentum carried away by the MHD disk wind to the specific angular momentum of the disk gas at the wind base (\citealt{Blandford82}, see also equations~(8) and (9) of \citealt{Tabone22a}). 
We set the dust mass loss rate $\dot{\Sigma}_{\rm d}^{\rm wind}$ to zero by assuming that the dust is settled below the wind base.

Turbulence induces not only disk accretion but also the diffusion and collisions of dust grains.
We parametrize the radial and vertical diffusion coefficients for gas, $D_{{\rm g},r}$, and $D_{{\rm g},z}$ as 
\beq
D_{{\rm g},r} = \alpha_{Dr} c_{\rm s}^2/\Omega, \qquad D_{{\rm g},z} = \alpha_{Dz} c_{\rm s}^2/\Omega,
\eeq
where $\alpha_{Dr}$ and $\alpha_{Dz}$ are the corresponding dimensionless diffusion coefficients.   
Assuming that the turbulence has a correlation time of $\approx 1/\Omega$, then
$\alpha_{Dr}c_{\rm s}^2$ and $\alpha_{Dz}c_{\rm s}^2$ stand for the  mean square radial and vertical velocity fluctuations caused by the turbulence, respectively \citep{FromangPapaloizou06}. We assume $\alpha_{\rm turb}$, and hence $D_{{\rm g},r}$ and $D_{{\rm g},r}$, to be vertically constant. Our default models consider nearly isotropic turbulence and take $\alpha_{Dr}$ and $\alpha_{Dz}$ to be
\beq
\alpha_{Dr} = \alpha_{Dz} = \frac{\alpha_{\rm turb}}{3}
\label{eq:alphaD}
\eeq
where $\alpha_{\rm turb}$ is the squared velocity dispersion normalized $c_{\rm s}^2$ (see \citealt{OkuzumiHirose11} for an example of MHD turbulence that indeed fulfills the relation $\alpha_{Dz} \approx  \alpha_{\rm turb}/3$). In section~\ref{sec:Dr}, we also consider the case where radial diffusivity is elevated due to a complex vertical structure of the radial flow around the midplane \citep{HuBai21}.

For simplicity, 
all the transport parameters $\alpha_{\rm visc}$, $\alpha_{\rm wind}$, $\lambda$, $\alpha_{\rm turb}$ are taken to be constant throughout the disk. If turbulence is the main driver of radial angular momentum transport, one expects $\alpha_{\rm turb} \sim \alpha_{\rm visc}$.

\subsection{Dust grains}\label{sec:grains}

To treat collisional grain size evolution at a low computational cost, we employ the single-size approach of \citet{Sato16}. In this approach, we assume that the dust mass budget at each radial position $r$ is dominated by grains of similar sizes, referred to as the mass-dominating grains. In many cases, these grains correspond to the largest grains at that position \citep{OrmalSpaans08,Sato16}. Hereafter, ``grains'' refer to these mass-dominating grains, unless otherwise noted. The radial distribution and size  of the grains is characterized by the surface mass density $\Sigma_{\rm d}$ and surface number density $N_{\rm d}$.  The mass of individual grains is related to these surface densities as $m_{\rm d} = \Sigma_{\rm d}/N_{\rm d}$. As demonstrated by \citet{Sato16}, this approach is valid for modeling the evolution of the dust mass budget. It should be noted that the ``grains'' are actually aggregates of smaller, (sub)micron-sized grains, which we call monomers. 

For simplicity, we approximate individual grains as spheres with radius $a = (3m_{\rm d}/(4\pi \rho_{\rm int}))^{1/3}$ and a fixed internal density $\rho_{\rm int}$. We compute the grain stopping time $t_{\rm stop}$ using equation (6) of \citet{Sato16}, which accounts for both Epstein and Stokes drag laws. In our simulations, Epstein's law applies to grains at $r \gtrsim 1~\rm au$. For these grains, the midplane Stokes number has a simple expression \citep[e.g.,][]{Birnstiel10a}
\beq
{\rm St}_{\rm mid} = \frac{\pi}{2}\frac{\rho_{\rm int}a}{\Sigma_{\rm g}}.
\label{eq:Stmid_Epstein}
\eeq

Assuming that the grains' vertical settling due to stellar gravity is balanced by vertical turbulent diffusion, we approximately have
\beq
\rho_{\rm d}(z) \approx \frac{\Sigma_{\rm d}}{\sqrt{2\pi}H_{\rm d}} \exp\left(-\frac{z^2}{2H_{\rm d}^2}\right),
\label{eq:rhod_Gaussian}
\eeq
where the dust scale height $H_{\rm d}$ can be written as \citep{Dubrulle95} 
\beq
H_{\rm d} = \left(1 + \frac{{\rm St}_{\rm mid}}{\alpha_{Dz}}\right)^{-1/2}H_{\rm g}.
\label{eq:Hd}
\eeq
Equation~\eqref{eq:rhod_Gaussian} provides a good approximation of $\rho_{\rm d}(z)$ at $z \lesssim H_{\rm g}$ (see equation~\eqref{eq:rhod_Takeuchi} in appendix~\ref{sec:local} for a more exact expression for $\rho_{\rm d}(z)$). 

Since we assume $D_{{\rm g},r}$  to be vertically constant, we can approximate the radial dust diffusion coefficient $D_{{\rm d}, r}$ by its value at the midplane, where most of the dust resides. We use the expression from \citet{Youdin07},
\beq
D_{{\rm d},r} = \frac{D_{{\rm g},r}}{1+{\rm St}_{\rm mid}^2}.
\eeq
For vertically constant $D_{{\rm d},r}$, the vertical integration of the diffusion term remaining in equation~\eqref{eq:Sigmad0} can be performed analytically. A detailed calculation of the vertical integration is provided in appendix~\ref{sec:diff}. Using the final result, given by equation~\eqref{eq:diffflux_final}, the evolutionary equation for $\Sigma_{\rm d}$ (equation~\eqref{eq:Sigmad0}) can be rewritten as
\beq
\frac{\pd\Sigma_{\rm d}}{\pd t} 
= -\frac{1}{r}\frac{\pd}{\pd r}r\left[
\brac{ v_{{\rm d},r}}_{\rm d}'\Sigma_{\rm d}
- \Sigma_{\rm g} D_{{\rm d},r}\frac{\pd}{\pd r}\pfrac{\Sigma_{\rm d}}{\Sigma_{\rm g}} 
\right],
\label{eq:Sigmad}
\eeq
where the effective advection velocity $\brac{ v_{{\rm d},r}}'$ is defined as
\beq
\brac{ v_{{\rm d},r}}_{\rm d}' \equiv \brac{ v_{{\rm d},r}}_{\rm d} -D_{{\rm d},r}\left(1-\pfrac{H_{\rm d}}{H_{\rm g}}^2\right)\frac{\pd\ln H_{\rm g}}{\pd r}.
\label{eq:vdravr_new}
\eeq
The second term of $\brac{ v_{{\rm d},r}}_{\rm d}'$ represents advection-like dust transport by diffusion, which occurs when the dust-to-gas mass ratio is vertically stratified, i.e., $H_{\rm d} < H_{\rm g}$  (see appendix~\ref{sec:diff} for an interpretation). Since $H_{\rm g}$ generally increases with $r$, this second term drives inward dust transport. This term was neglected in previous 1D models for radial dust transport in the literature, but it can contribute significantly to inward dust transport when $D_{{\rm d},r}$ is high (see section~\ref{sec:Dr}).

The evolutionary equation for $N_{\rm d}$ is given by
\beq
\frac{\pd N_{\rm d}}{\pd t} 
= -\frac{1}{r}\frac{\pd}{\pd r}r\left[
\brac{ v_{{\rm d},r}}' N_{\rm d}
- \Sigma_{\rm g} D_{{\rm d},r}\frac{\pd}{\pd r}
\pfrac{N_{\rm d}}{\Sigma_{\rm g}}
\right] 
- \xi_{\rm stick} \frac{N_{\rm d}}{t_{\rm coll}},
\label{eq:Nd}
\eeq
where $t_{\rm coll}$ is the mean collision time of the grains and $\xi_{\rm stick}$ is a dimensionless coefficient introduced to account for grain fragmentation  at high collision speeds. As in equation~\eqref{eq:Sigmad}, the two terms in the brackets in equation~\eqref{eq:Nd} represent the radial advection and diffusion fluxes. The last term represents the decrease or increase in $N_{\rm d}$ due to collisional growth or fragmentation.
The mean collision time $t_{\rm coll}$ depends on the vertical distribution of the dust density.
Using equation~\eqref{eq:rhod_Gaussian}, we approximately have \citep{Sato16} 
\begin{equation}
t_\mathrm{coll}
\approx \frac{H_{\rm d}}{2\sqrt{\pi} a^2 \Delta v N_{\rm d}},
\end{equation}
where $\Delta v$ is the grain collision velocity averaged over the grain vertical distribution. 
The collision velocity accounts for Brownian motion, radial and azimuthal drift, vertical settling, and turbulence, with turbulence strength given by $\alpha_{\rm turb}$ \citep{Sato16}.  
Following \citet{Sato16}, we assume a typical Stokes number ratio of 0.5 for the colliding grains.

The sticking coefficient $\xi_{\rm stick}$ can be either positive or negative depending on whether the grains gain or lose mass upon collision. 
Following \citet{OH12} and \citet{Okuzumi16}, we model $\xi_{\rm stick}$ as
\beq
\xi_{\rm stick}=
    {\rm min} \left\{1,-\frac{\ln(\Delta v/v_{\rm stick})}{\ln 5}\right\} ,
\eeq
where $v_{\rm stick}$ is the threshold velocity below which colliding grains can grow in mass.
As discussed in section~\ref{sec:intro}, the threshold sticking velocity is highly uncertain, so we treat $v_{\rm stick}$ as a constant free parameter (see also section~\ref{sec:runs}). 

\subsection{Accretion flow models}
\label{sec:vertical}
To demonstrate the impact of a vertically varying accretion flow on dust transport, we consider two disk models in which the wind-driven accretion flow is either vertically uniform or narrowly concentrated near the base of the MHD wind (see figure~\ref{fig:flow} and section~\ref{sec:examples}). We refer to the former and latter as the uniform accretion model and surface accretion model, respectively. 
For simplicity, both models assume vertically uniform viscosity-driven flow, i.e.,  $v_{{\rm g},r}^{\rm visc}(z) = \brac{v_{{\rm g},r}^{\rm visc}}_{\rm g}$ at all $z$.

For the uniform accretion model, the net dust radial velocity $\brac{ v_{{\rm d},r}}_{\rm d}$ (equation~\eqref{eq:vdravr_approx}) is given by (see equation~\eqref{eq:coacc_uniform})
\beq
\brac{ v_{{\rm d},r}}_{\rm d} =
\frac{ \brac{v_{{\rm g},r}^{\rm visc}}_{\rm g} + \brac{v_{{\rm g},r}^{\rm wind}}_{\rm g} }{1+{\rm St}_{\rm mid}^2}
+ \frac{2 {\rm St}_{\rm mid}\Delta v_{{\rm g},\phi,{\rm mid}}}{1+{\rm St}_{\rm mid}^2}.
\label{eq:vdravr_uniform}
\eeq
For the surface accretion model, we neglect co-accretion of dust with the wind-driven accretion by setting $C_{\rm surface} = 0$ throughout the disk. This simplification can be generally justified if ${\rm St}_{\rm mid} > \alpha_{Dz}$ (see appendix~\ref{sec:local}), which is satisfied in our simulations (see section~\ref{sec:results}).
The net dust radial velocity for the surface accretion model is thus given by 
\beq
\brac{ v_{{\rm d},r}}_{\rm d} = 
\frac{ \brac{v_{{\rm g},r}^{\rm visc}}_{\rm g} }{1+{\rm St}_{\rm mid}^2}
+ \frac{2 {\rm St}_{\rm mid}\Delta v_{{\rm g},\phi,{\rm mid}}}{1+{\rm St}_{\rm mid}^2}.
\label{eq:vdravr_surface}
\eeq

In section~\ref{sec:formulation}, we predicted that dust retention will occur when equation~\eqref{eq:cond_dustretention} is met. We now rewrite this condition for the surface accretion model. Assuming ${\rm St}_{\rm mid} \ll 1$,  $\brac{ v_{{\rm d},r}}_{\rm d}$ can be approximated as
\beq
\brac{ v_{{\rm d},r}}_{\rm d} \approx 
\brac{v_{{\rm g},r}^{\rm visc}}_{\rm g}
+ {2 {\rm St}_{\rm mid}\Delta v_{{\rm g},\phi,{\rm mid}}},
\eeq
Using this and $\brac{ v_{{\rm g},r}}_{\rm g} = \brac{v_{{\rm g},r}^{\rm visc}}_{\rm g} + \brac{v_{{\rm g},r}^{\rm wind}}_{\rm g}$,  equation~\eqref{eq:cond_dustretention} can be rewritten as  (cf. equation~\eqref{eq:St_dustretention}) 
\beq
{\rm St}_{\rm mid} < \frac{\brac{v_{{\rm g},r}^{\rm wind}}_{\rm g}}{2\Delta v_{{\rm g},\phi,{\rm mid}}}.
\label{eq:St_dustretention_model}
\eeq
This criterion does not involve $\brac{v_{{\rm g},r}^{\rm visc}}_{\rm g}$ since both dust and gas at the midplane accrete at this velocity. However,  this is a necessary but not sufficient condition for dust retention: achieving appreciable dust retention  additionally requires that $|\brac{v_{{\rm g},r}^{\rm visc}}_{\rm g}| < |\brac{v_{{\rm g},r}^{\rm wind}}_{\rm g}|$, or equivalently, $\alpha_{\rm visc} < \alpha_{\rm wind}$.

If we further use equations~\eqref{eq:dvgphi} and \eqref{eq:vgr_wind}, equation~\eqref{eq:St_dustretention_model} simplifies to
\beq
{\rm St}_{\rm mid} < \frac{3\alpha_{\rm wind}}{|\pd\ln P/\pd\ln r|_{\rm mid}}.
\label{eq:St_dustretention_simple}
\eeq
If $(\partial \ln P/\partial\ln r)_{\rm mid} \sim -3$, which approximately holds in the inner part of our disk models (see section~\ref{sec:default}), equation~\eqref{eq:St_dustretention_simple} reduces to ${\rm St}_{\rm mid} \lesssim \alpha_{\rm wind}$. 
In the following section, we test whether dust retention indeed occurs when equation~\eqref{eq:St_dustretention_simple} is met.




\subsection{Numerics, initial conditions, and parameter choices}
\label{sec:runs}
We numerically solve the conservation equations~\eqref{eq:Sigmag}, \eqref{eq:Sigmad}, and \eqref{eq:Nd} using an explicit-in-time finite volume scheme. The computational domain spans $0.3~{\rm au} \leq r \leq 300~{\rm au}$ and is divided into 300 logarithmically spaced cells. We treat the first and second terms in the brackets in the conservation equations as the advection and diffusion fluxes, respectively, and compute the fluxes at the cell boundaries using a first-order upwind scheme for advection and a central difference scheme for diffusion.
At the inner boundary of the computational domain, we impose outflow and zero-flux boundary conditions for the advection and diffusion fluxes, respectively.
At the outer boundary, we impose a zero-flux boundary condition for both the advection and diffusion fluxes.

The initial gas surface density distribution, $\Sigma_{\rm g,0}$, is given by a power law tapered by an exponential cutoff,
\beq
\Sigma_{\rm g,0}(r) = \frac{M_{\rm disk,0}}{2\pi r_{\rm c,0}^2\Gamma(1+\xi)}\pfrac{r}{r_{\rm c,0}}^{-1+\xi}\exp\left(-\frac{r}{r_{\rm c,0}}\right),
\label{eq:Sigmag0}
\eeq
where $\Gamma$ is the Gamma function, $M_{\rm disk,0}$ and $r_{\rm c,0}$ are the initial disk mass and initial characteristic radius, and $\xi = (2(\lambda-1))^{-1} \alpha_{\rm wind}/(\alpha_{\rm visc}+\alpha_{\rm wind})$.
Equation~\eqref{eq:Sigmag0} represents the self-similar solution to equation~\eqref{eq:Sigmag}, assuming $T \propto r^{-1/2}$, with radially constant $\alpha_{\rm visc}$ and $\alpha_{\rm wind}$ \citep{Tabone22a}. The initial dust surface density and grain size are set to $\Sigma_{\rm d} = 0.01\Sigma_{\rm g}$ and $a = 0.1~\rm \mu m$, respectively.

\begin{table}[t] 
    \tbl{Summary of default model parameters}{
    \begin{tabular}{llc}
    \hline 
    Symbol & Description & Values \\
    \hline
     $M_*$  & Stellar mass & $1M_\odot$\\    
     $M_{{\rm disk},0}$ & Initial disk mass & $0.1M_\odot$ \\     $r_{\rm c,0}$ & Initial disk characteristic radius & 30 au \\   
     $\alpha_{\rm visc}$  & Viscosity parameter & $3\times 10^{-4}$\\
     $\alpha_{\rm wind}$  & Wind stress parameter & $6\times 10^{-3}$\\     
     $\lambda$ & Lever arm parameter & 3 \\
     $\alpha_{\rm turb}$  & Turbulence strength parameter & $=\alpha_{\rm visc}$\\  
     $\alpha_{Dr}$ & Radial diffusion  parameter & $=\alpha_{\rm turb}/3$\\   
     $\alpha_{Dz}$  & Vertical diffusion  parameter & $=\alpha_{\rm turb}/3$\\        
     $\rho_{\rm int}$ &  Grain internal density & $0.6~{\rm g~cm^{-3}}$  \\
     $v_{\rm stick}$ &  Grain sticking threshold velocity & $\{0.1, 0.3, 1\}~{\rm m~s^{-1}}$ \\ 
    \hline
    \end{tabular}
    }
    \label{tab:param}    
\end{table}

Table~\ref{tab:param} summarizes the default values of the key model parameters. The initial disk mass is set such that the disk is initially barely gravitationally stable. Our choice for $r_{\rm c,0}$ falls within the range of the characteristic radii of the disks in the $\sim 1$ Myr-old Lupus and Taurus star-forming regions ($r_{\rm c} \sim 10$--$70~\rm au$) indirectly estimated by \citet{Trapman23} based on observed CO gas disk radii\footnote{Note that the apparent CO gas disk radii are likely much larger than $r_{\rm c}$ because CO emission is optically thick \citep{Trapman19,Trapman23}.}. This also aligns with their dust disk radii ($\sim 20$--$100~\rm au$) from millimeter dust continuum observations \citep{Hendler20}.

The choice of $\lambda$ follows \citet{Tabone22a} and \citet{Trapman22}. With this choice, the wind mass loss rate is comparable to the wind-driven mass accretion rate, which is broadly consistent with observational constraints on MHD disk winds \citep{Tabone22a}. The adopted wind stress parameter, $\alpha_{\rm wind} = 6\times 10^{-3}$, results in an initial gas accretion timescale of $t_{\rm adv,g}(r_{\rm c}) \sim r_{\rm c}/|\brac{v_{{\rm g},r}^{\rm wind}}_{\rm g}(r_{\rm c})| \approx 0.9$ Myr. With these default parameters, the mass accretion rate $-2\pi r \brac{v_{{\rm g},r}}_{\rm g}\Sigma_{\rm g}$ measured at the inner computational boundary is $\sim 10^{-8} M_\sun~\rm yr^{-1}$ at $t \sim 2$ Myr and $\sim 10^{-9} M_\sun~\rm yr^{-1}$ at $t \sim 5$ Myr, which is consistent with the median accretion rates of solar-mass young stars in the Lupus and Upper Scorpius regions, respectively \citep[][see their figure 11]{Testi22}.
We note that recent MHD simulations of wind-driven accretion \citep{Bethune17,Bai17,Iwasaki24} tend to predict smaller lever arm parameters of $\lambda \sim 1.5$. However, using $\lambda = 1.5$ results in the accretion rate at our inner computational boundary being an order of magnitude lower than that for $\lambda = 3$, making it too small to be consistent with the observed accretion rates.

Motivated by the recent studies introduced in section~\ref{sec:intro}, we assume fragile aggregates and adopt a default value of $v_{\rm stick} = 0.3~\rm m~s^{-1}$. We also consider more and less sticky cases with $v_{\rm stick} = 0.1$ and $1~\rm m~s^{-1}$, respectively. These values are significantly lower than the conventionally assumed values of $v_{\rm stick} = 10$--$70~\rm m~s^{-1}$ for water ice \citep[e.g.,][]{Wada09,Wada13,Gundlach15}, but are closer to the values for water and CO$_2$ ice suggested by more recent experiments and observations \citep{Musiolik16a,Musiolik16b,Gundlach18,Okuzumi19,Jiang24,Ueda24}.

Our choice of $\rho_{\rm int}$ assumes moderately porous aggregates of ice and dust \citep{Okuzumi19}. Varying $\rho_{\rm int}$ has little effect on the simulated dust evolution as long as the grains obey Epstein's drag law \citep[e.g.,][]{Okuzumi12}.

\section{Simulation results}\label{sec:results}

\subsection{Uniform versus surface accretion}\label{sec:default}

We begin by presenting the simulation results from the uniform and surface accretion models with $v_{\rm stick} = 0.3~\rm m~s^{-1}$, focusing on how the vertical structure of the gas accretion flow affects global dust transport.

\begin{figure*}
\begin{center}
\includegraphics[width=\hsize, bb=0 0 595 387]{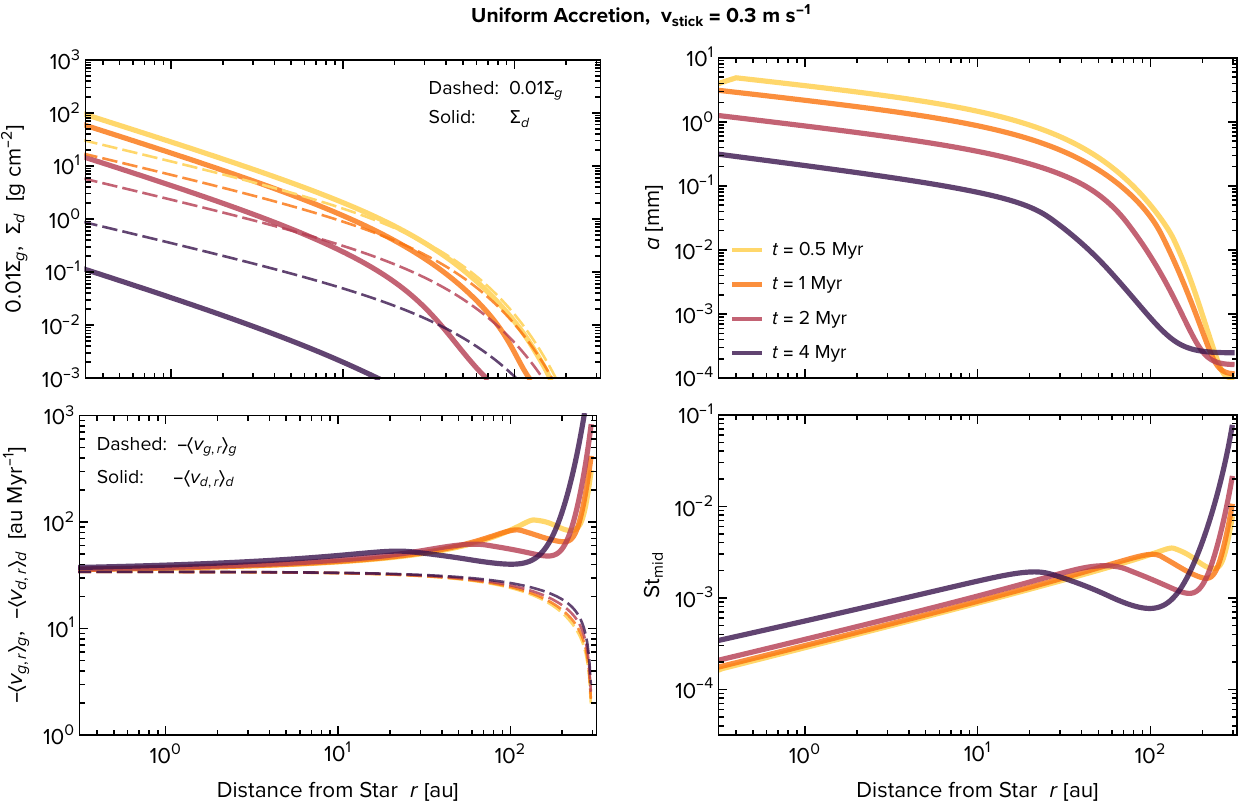}
\end{center}
\caption{Gas and dust evolution from the uniform accretion disk model with $v_{\rm stick} = 0.3~\rm m~s^{-1}$. The upper and lower left panels show the radial distribution of the surface densities and density-weighted average accretion velocities, respectively, for the gas and dust at different times $t$. The upper and lower right panels are for the size and midplane Stokes number of the mass-dominating grains. 
}
\label{fig:gasdust_uniform}
\end{figure*}

\begin{figure*}
\begin{center}
\includegraphics[width=\hsize, bb=0 0 595 387]{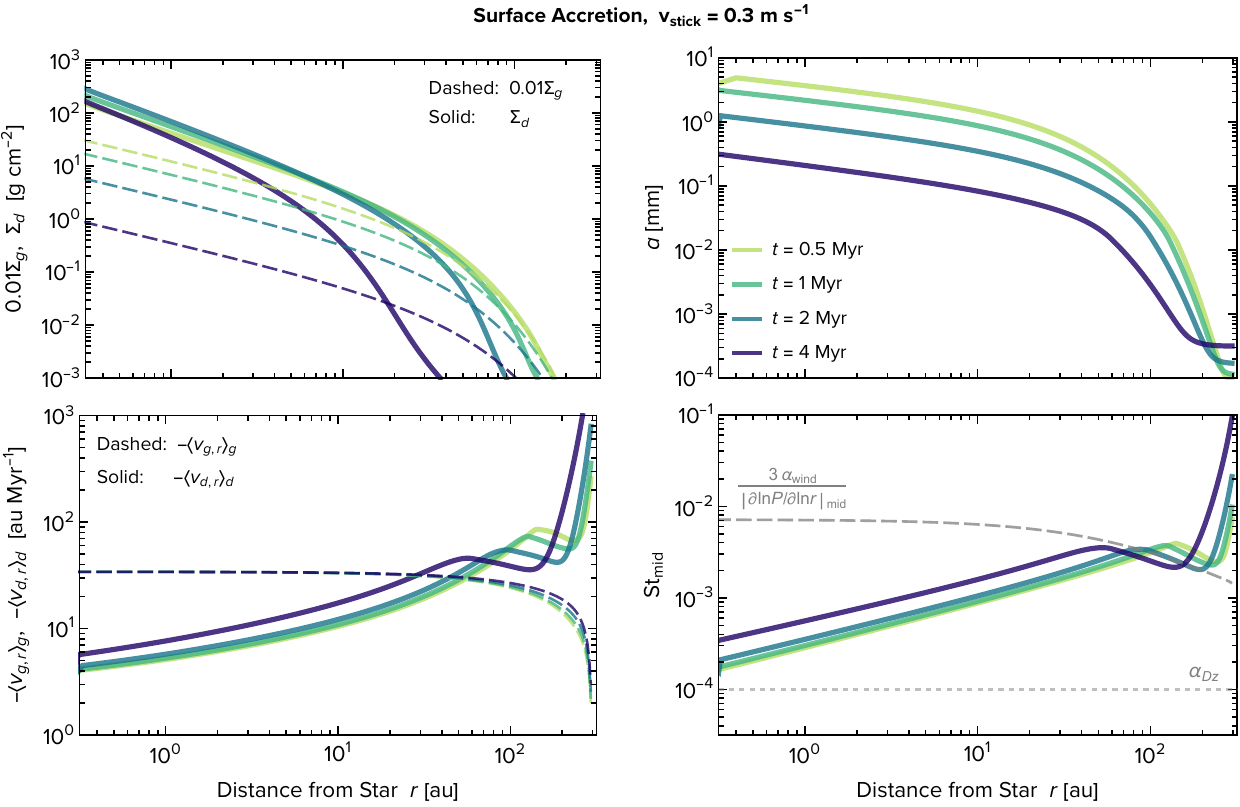}
\end{center}
\caption{Same as figure~\ref{fig:gasdust_uniform}, with $v_{\rm stick} = 0.3~\rm m~s^{-1}$. In the lower right panel, the gray dashed and dotted lines mark ${\rm St}_{\rm mid} = 3\alpha_{\rm wind}/|\pd\ln P/\pd\ln r|_{\rm mid}$ and ${\rm St}_{\rm mid} = \alpha_{Dz}$ respectively. The pressure gradient slope hardly depends on $t$, so only the dashed line for $t = 1~\rm Myr$ is shown. Dust retention occurs if ${\rm St}_{\rm mid}$ lies below the dashed line (equation~\eqref{eq:St_dustretention_simple}), while the assumption that $C_{\rm surface} \approx 0$ can be safely justified if ${\rm St}_{\rm mid}$ lies above the dotted line (see figure~\ref{fig:Csurface} in appendix~\ref{sec:local}).

}
\label{fig:gasdust_surface}
\end{figure*}

Figures~\ref{fig:gasdust_uniform} and \ref{fig:gasdust_surface} show the gas and dust evolution obtained from the two models. Since both models adopt the same viscosity and wind parameters, they produce identical radial profiles of the gas transport velocity $\brac{v_{\rm g,r}}_{\rm g}$ (lower left panels). As a result, the evolution of the gas surface density $\Sigma_{\rm g}$ (upper left panels) is identical in these models. They also produce nearly identical results for the evolution of the size $a$ and midplane Stokes numbers ${\rm St}_{\rm mid}$ of the mass-dominating dust grains (upper and lower right panels, respectively). In both models, grains at $r \lesssim 100~\rm au$ grow until their size reaches the limit set by collisional fragmentation\footnote{At $r \gtrsim 100~\rm au$, radial drift rather than fragmentation limits local dust growth.}. When turbulence is the main driver of the grains' relative velocity, the maximum Stokes number set by collisional fragmentation can be estimated as \citep{Birnstiel09,Okuzumi19}
\begin{align}
{\rm St}_{\rm mid, frag} &= \frac{ v_{\rm stick}^2}{2.3\alpha_{\rm turb}c_{\rm s}^2}
\notag \\
&\approx 1\times10^{-3} \pfrac{v_{\rm stick}}{0.3~\rm m~s^{-1}}^{2}\pfrac{\alpha_{\rm turb}}{3\times 10^{-4}}^{-1}
\pfrac{T}{30~\rm K}^{-1},
\label{eq:Stfrag}
\end{align}
where the reference temperature of 30 K corresponds to the temperature at $r \sim r_{\rm c} = 30~\rm au$ in our model. Using Epstein's drag law (equation~\eqref{eq:Stmid_Epstein}), this maximum Stokes number translates into a maximum grain size of
\begin{align}
a_{\rm frag} 
&\approx \frac{0.3 v_{\rm stick}^2 \Sigma_{\rm g}}{\alpha_{\rm turb} c_{\rm s}^2\rho_{\rm int}}
\nonumber \\
&\approx 0.1 \pfrac{v_{\rm stick}}{0.3~\rm m~s^{-1}}^{2} \pfrac{\alpha_{\rm turb}}{3 \times 10^{-4}}^{-1} \pfrac{\rho_{\rm int}}{0.6 ~\rm g~cm^{-3}}^{-1} 
\nonumber \\
& \quad \times  \pfrac{\Sigma_{\rm g}}{10~\rm g~cm^{-2}}\pfrac{T}{30~\rm K}^{-1} ~\rm mm.
\label{eq:afrag}
\end{align}
Equations~\eqref{eq:Stfrag} and \eqref{eq:afrag} reproduce the simulation results shown in figures~\ref{fig:gasdust_uniform} and \ref{fig:gasdust_surface}. These estimates do not depend on the vertical distribution of the accretion flow,  explaining why the two models produce nearly identical results for ${\rm St}_{\rm mid}$ and $a$.

The uniform and surface accretion models, however, predict very different evolution for the dust surface density, as shown in the upper left panels of figures~\ref{fig:gasdust_uniform} and \ref{fig:gasdust_surface}. We find that $\Sigma_{\rm d}$ decreases faster than $\Sigma_{\rm g}$ in the uniform accretion model, while the opposite occurs in the surface accretion model. These results are consistent with the predictions made in section~\ref{sec:examples}. In the uniform accretion model, the dust transport velocity $|\langle v_{{\rm d},r}\rangle_{\rm d}|$ never falls below the gas transport velocity $|\langle v_{{\rm g},r}\rangle_{\rm g}|$ (see the lower left panel of figure~\ref{fig:gasdust_uniform}), resulting in  $t_{\rm adv,d} < t_{\rm adv,g}$ everywhere.  In contrast, the surface accretion model results in $|\langle v_{{\rm d},r}\rangle_{\rm d}| < |\langle v_{{\rm g},r}\rangle_{\rm g}|$ at $r \lesssim 50~\rm au$ (see the lower left panel of figure~\ref{fig:gasdust_surface}), indicating that the condition $t_{\rm adv, d} > t_{\rm adv,g}$ for dust retention is met in that region. As a further check, the lower right panel of figure~\ref{fig:gasdust_surface}  confirms that ${\rm St}_{\rm mid}$ in the surface accretion model fulfills the criterion for dust retention, equation~\eqref{eq:St_dustretention_simple},  at $r \lesssim 50~\rm au$. {At $r \lesssim 10~\rm au$, the gas pressure has a constant radial slope $\pd \ln P/\pd\ln r \approx -2.5$, and hence equation~\eqref{eq:St_dustretention_simple} reduces to ${\rm St}_{\rm mid} < 1.2\alpha_{\rm wind}$.}

\begin{figure*}[t]
\begin{center}
\includegraphics[width=0.48\hsize, bb=0 0 289 263]{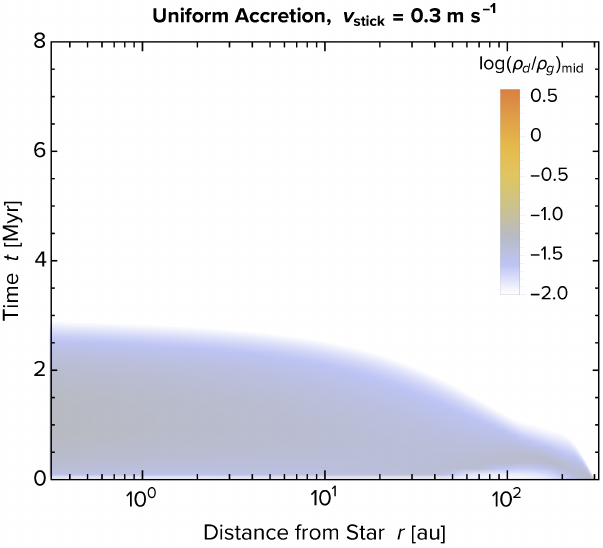}
\hspace{4mm}
\includegraphics[width=0.48\hsize, bb=0 0 289 263]{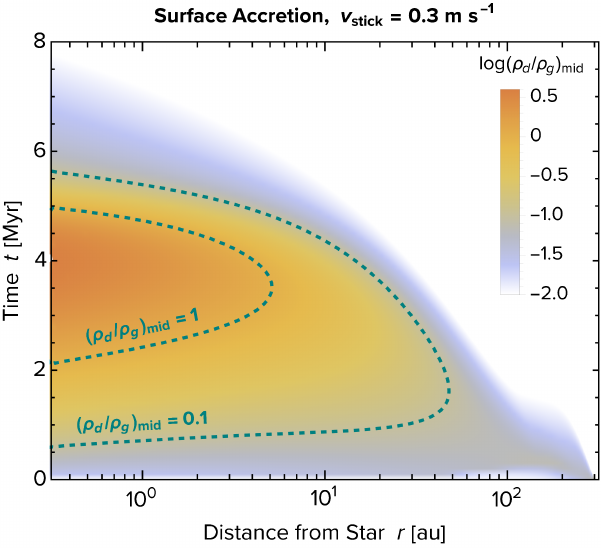}
\end{center}
\caption{Midplane dust-to-gas mass ratio $(\rho_{\rm d}/\rho_{\rm g})_{\rm mid} = (\Sigma_{\rm d}/\Sigma_{\rm g})/(H_{\rm d}/H_{\rm g})$ as a function of orbital radius $r$ and time $t$ from the uniform and surface accretion models (left and right panels, respectively) with $v_{\rm stick} = 0.3~\rm m~s^{-1}$.
}
\label{fig:fdg}
\end{figure*}
{Simulations of the streaming instability with and without externally driven turbulence show that} strong dust clumping, which leads to planetesimal formation, occurs when the midplane dust-to-gas mass ratio $(\rho_{\rm d}/\rho_{\rm g})_{\rm mid}$ exceeds $\sim O(1)$ {\citep{Gole20,LiYoudin21}. To examine whether this condition can be achieved in the two disk models considered here, we plot} $(\rho_{\rm d}/\rho_{\rm g})_{\rm mid} = (\Sigma_{\rm d}/\Sigma_{\rm g})/(H_{\rm d}/H_{\rm g})$ from these models as a function of $r$ and $t$ in Figure~\ref{fig:fdg}. In both models, $(\rho_{\rm d}/\rho_{\rm g})_{\rm mid}$ initially increases from 0.01 to $\sim$ 0.03 due to local dust growth and subsequent settling. In the uniform accretion model, no further increase of $(\rho_{\rm d}/\rho_{\rm g})_{\rm mid}$ occurs because dust and gas accrete at similar speeds ($|\langle v_{{\rm d},r}\rangle_{\rm d}| \approx |\langle v_{{\rm g},r}\rangle_{\rm g}|$), as shown in the lower left panel of figure~\ref{fig:gasdust_uniform}. In contrast, in the surface accretion model, $(\rho_{\rm d}/\rho_{\rm g})_{\rm mid}$ increases further, exceeding unity at $t \sim$ 2--3 Myr. The region with $(\rho_{\rm d}/\rho_{\rm g})_{\rm mid} > 1$ extends out to $r \approx 10$ au at its maximum and persists until $t \sim 4$ Myr, after which the grains are lost due to radial drift. 

As noted in section~\ref{sec:vertical}, the assumption $C_{\rm surface} \approx 0$ made in the surface accretion model is valid as long as ${\rm St}_{\rm mid} > \alpha_{ Dz}$. The lower right panel of figure~\ref{fig:gasdust_surface} shows that the simulation presented here indeed satisfies this condition.

In the surface accretion model presented here, radial dust diffusion has little effect on dust retention. This is because the radial diffusion timescale $\sim r^2/D_{{\rm g},r} \sim (r/H_{\rm g})^2/(\alpha_{Dr}\Omega)$ is longer than the local dust advection timescale $t_{\rm adv,d} \sim r/|2{\rm St}_{\rm mid}\Delta v_{\rm g,\phi,mid}| \sim (r/H_{\rm g})^2/({\rm St}_{\rm mid}\Omega)$ when  ${\rm St}_{\rm mid} > \alpha_{Dr}$. This condition is satisfied in this default model, where $\alpha_{Dr} = \alpha_{Dz}$ and ${\rm St}_{\rm mid} > \alpha_{ Dz}$. However, radial dust diffusion can become critical when $\alpha_{Dr} \gg \alpha_{Dz}$, as we show in section~\ref{sec:Dr},.

\subsection{How small must $v_{\rm stick}$ be for dust retention?}
\label{sec:vstick}

Since ${\rm St}_{\rm frag}$ scales quadratically with $v_{\rm stick}$ (see equation~\eqref{eq:Stfrag}), the threshold sticking velocity $v_{\rm stick}$ critically affects whether the criterion for dust retention (equation~\eqref{eq:St_dustretention_simple}) can be fulfilled. Specifically, the combination of   equations~\eqref{eq:St_dustretention_simple} and \eqref{eq:Stfrag} predicts that dust retention requires 
\begin{align}
v_{\rm stick} &\lesssim \sqrt{2\alpha_{\rm turb}\alpha_{\rm wind}} c_{\rm s} \notag \\
& \approx 0.6
\pfrac{\alpha_{\rm turb}}{3\times10^{-4}}^{1/2}
\pfrac{\alpha_{\rm wind}}{6\times10^{-3}}^{1/2}
\pfrac{T}{30~\rm K}^{1/2}~\rm m~s^{-1}.
\label{eq:vstick_dustretention}
\end{align}

\begin{figure*}[t]
\begin{center}
\includegraphics[width=0.48\hsize, bb=0 0 289 263]{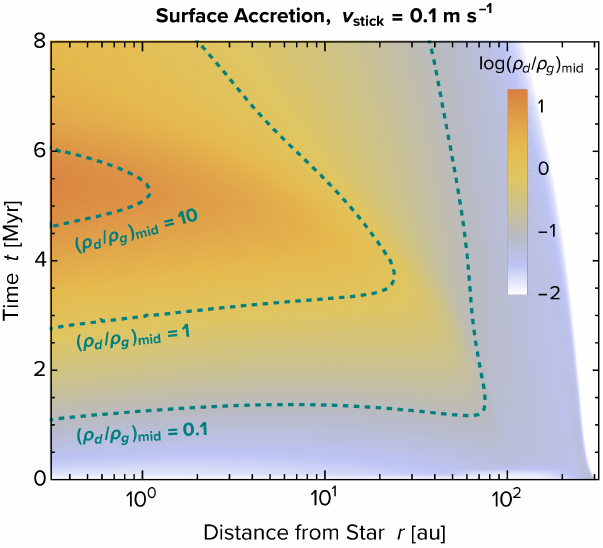}
\hspace{4mm}
\includegraphics[width=0.48\hsize, bb=0 0 289 263]{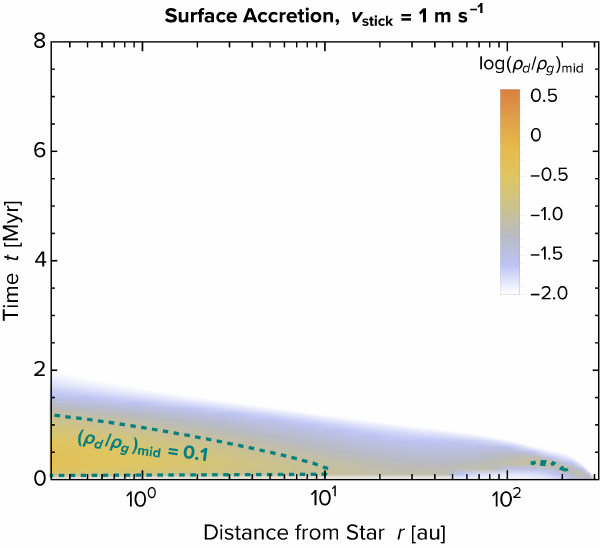}
\end{center}
\caption{Same as figure~\ref{fig:fdg}, but from the surface accretion models with $v_{\rm stick} = 0.1$ and 1 $\rm m~s^{-1}$ (left and right panels, respectively). Note that the colors in the left and right panels indicate different values.
}
\label{fig:fdg_vstick}
\end{figure*}
To test this prediction, we show in figure~\ref{fig:fdg_vstick} the space--time plots of $(\rho_{\rm d}/\rho_{\rm g})_{\rm mid}$ for surface accretion models with $v_{\rm stick} = 0.1$ and $1~\rm m~s^{-1}$. As expected, the degree of dust retention depends critically on $v_{\rm stick}$. For $v_{\rm stick} = 0.1~\rm m~s^{-1}$,  $(\rho_{\rm d}/\rho_{\rm g})_{\rm mid}$ exceeds unity at wider orbits and over longer timescales than for $v_{\rm stick} = 0.3~\rm m~s^{-1}$ (see the right panel of figure~\ref{fig:fdg}).
In contrast, for $v_{\rm stick} = 1~\rm m~s^{-1}$, $(\rho_{\rm d}/\rho_{\rm g})_{\rm mid}$ never exceeds unity within our computational domain ($r > 0.3~\rm au$). This result is broadly consistent with equation~\eqref{eq:vstick_dustretention}, which requires $v_{\rm stick} \lesssim 1~\rm m~s^{-1}$ for dust retention in this domain.
We note that the sufficient condition ${\rm St}_{\rm mid} > \alpha_{ Dz}$ for $C_{\rm surface} \approx 0$ is satisfied at $r \gtrsim 1~\rm au$ even in the $v_{\rm stick} = 1~\rm m~s^{-1}$ model.

\begin{figure}[t]
\begin{center}
\includegraphics[width=\hsize, bb=0 0 289 263]{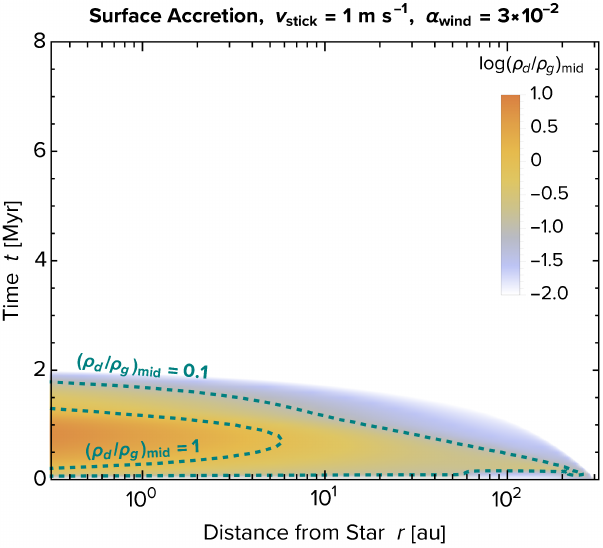}
\end{center}
\caption{Same as figure~\ref{fig:fdg}, but from the surface accretion model with $v_{\rm stick} = 1~\rm m~s^{-1}$ and with an elevated wind stress parameter of  $\alpha_{\rm wind} = 3\times 10^{-2}$. Note that the colors in this figure and figure~\ref{fig:fdg} indicate slightly different values.
}
\label{fig:fdg_vstick_alphawind}
\end{figure}
Equation~\eqref{eq:vstick_dustretention} indicates that the viability of dust retention also depends on both $\alpha_{\rm turb}$ and $\alpha_{\rm wind}$. 
Specifically, a higher $\alpha_{\rm wind}$ allows dust retention with a higher $v_{\rm stick}$. We demonstrate this in figure~\ref{fig:fdg_vstick_alphawind}, where we show the midplane dust-to-gas mass ratio from the surface accretion model with $v_{\rm stick} = 1~\rm m~s^{-1}$ and  an elevated wind stress parameter of $\alpha_{\rm wind} = 3\times 10^{-2}$. 
In contrast to the default $v_{\rm stick} = 1~\rm m~s^{-1}$ model with $\alpha_{\rm wind} = 6\times 10^{-3}$ shown in the right panel of figure~\ref{fig:fdg_vstick},  $(\rho_{\rm d}/\rho_{\rm g})_{\rm mid}$ in this model exceeds unity, consistent with the prediction from equation~\eqref{eq:vstick_dustretention}. The duration of dust retention is shorter than in the default model due to the faster wind-driven accretion and radial drift.

At first glance, equation~\eqref{eq:vstick_dustretention} seems to suggest that dust retention favors larger $\alpha_{\rm turb}$. This is because stronger turbulence leads to a smaller ${\rm St}_{\rm mid}$ and, consequently, slower inward drift. However, dust retention does not always favor strong turbulence. As discussed in section~\ref{sec:vertical}, a significant level of dust retention requires $\alpha_{\rm visc} < \alpha_{\rm wind}$ (i.e., surface accretion dominates the vertically integrated gas accretion) in addition to  equation~\eqref{eq:vstick_dustretention}. Since we assume $\alpha_{\rm visc} = \alpha_{\rm diff}$, this additional requirement is equivalent to $\alpha_{\rm turb} < \alpha_{\rm wind}$.

{We note that whether surface accretion leads to dust retention is insensitive to the details of the assumed dust coagulation model. The criterion for dust retention, equation~\eqref{eq:St_dustretention_simple}, depends only on the Stokes number and is independent of other grain properties\footnote{We ran a test simulation with grains of fixed Stokes number ${\rm St}_{\rm mid} = 10^{-3}$ and confirmed that dust retention occurs.}. While our model accounts only for coagulation and fragmentation, bouncing can also limit the grain Stokes number  \citep[e.g.,][]{Guttler10,Zsom11,DominikDullemond24}. In any case, dust retention will occur when equation~\eqref{eq:St_dustretention_simple} is fulfilled.}

\subsection{Effects of {anisotropic} dust diffusion} 
\label{sec:Dr}
Here, we study how our simulation results change when we relax the assumption that $\alpha_{Dr} = \alpha_{Dz}$. \citet{HuBai21} demonstrated that complex midplane flows produced by MHD, combined with turbulent vertical diffusion, can lead to effective radial gas diffusion. For the vertical diffusivity $\alpha_{Dz} = 1\times 10^{-4}$ adopted in our model, \citet{HuBai21} found that the effective radial diffusivity $\alpha_{Dr}$ for grains with ${\rm St}_{\rm mid} \sim 10^{-4}$--$10^{-3}$ can be enhanced to $\sim 10^{-3}$, with the exact value depending on the details of the midplane flow structure (see their figures 5 and 6). Motivated by this, we reran the default surface accretion simulation presented in section~\ref{sec:default}, but with elevated radial dust diffusivities of $\alpha_{Dr} = 10^{-3}$ and $10^{-2}$. 

\begin{figure*}[t]
\begin{center}
\includegraphics[width=0.48\hsize, bb=0 0 289 263]{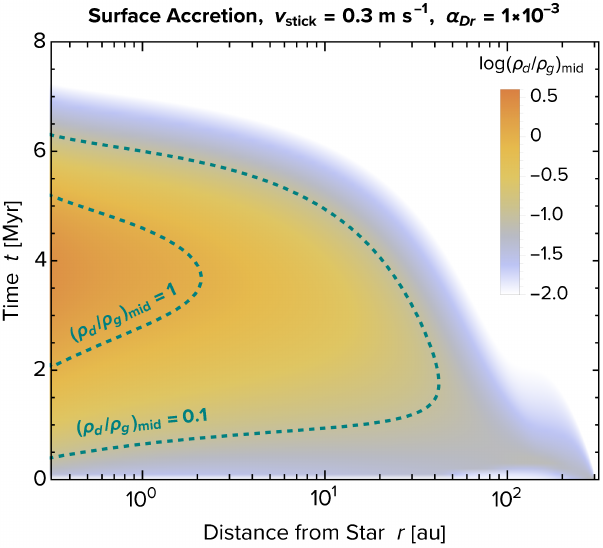}
\hspace{4mm}
\includegraphics[width=0.48\hsize, bb=0 0 289 263]{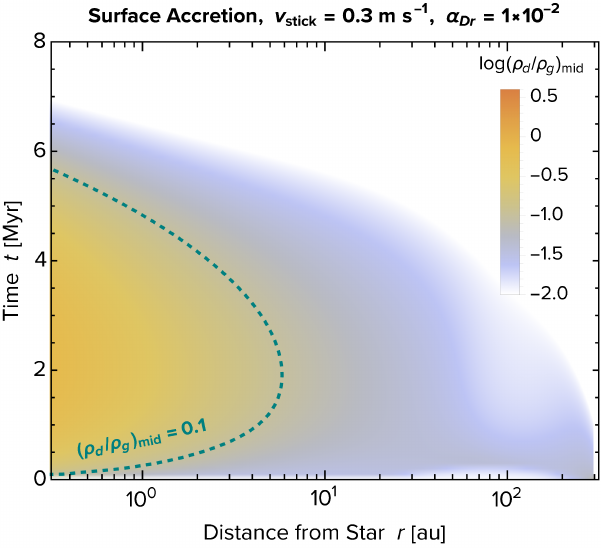}
\end{center}
\caption{Same as figure~\ref{fig:fdg}, but from the surface accretion models with elevated radial diffusivities of $\alpha_{Dr} = 10^{-3}$ and $10^{-2}$ (left and right panels, respectively).
}
\label{fig:fdg_Ddr}
\end{figure*}
The results are presented in figure~\ref{fig:fdg_Ddr}. Overall, we find that enhancing radial diffusivity reduces the efficiency of dust retention. This reduction is caused by the effective inward advection identified in appendix~\ref{sec:diff},  represented by the second term in equation~\eqref{eq:vdravr_new}\footnote{In contrast, the standard radial diffusion flux $-D_{{\rm g},r}\Sigma_{\rm g} \pd(\Sigma_{\rm d}/\Sigma_{\rm g})/\pd r$ is outward since $\Sigma_{\rm d}/\Sigma_{\rm g}$ has a negative radial gradient (see the upper left panel of figure~\ref{fig:gasdust_surface}).}. However, this effect becomes significant only when $\alpha_{Dr} = 10^{-2}$; for $\alpha_{Dr} = 10^{-3}$, the effect is relatively minor, with the midplane dust-to-gas ratio still exceeding unity. We thus conclude that the radial dust mixing in complex midplane MHD flows, as identified by \citet{HuBai21}, would have only a minor effect on the dust retention mechanism proposed in this work.

{The opposite situation ($\alpha_{Dr} < \alpha_{Dz}$) can occur when the disk has vertically elongated turbulence driven by the vertical shear instability \citep[e.g.,][]{StollKley16}. In this case, radial dust diffusion is always negligible (${\rm St}_{\rm mid} > \alpha_{Dr}$; see the last paragraph of Section \ref{sec:default}) as long as the dust has grown sufficiently to settle below the surface accretion layer (${\rm St}_{\rm mid} > \alpha_{Dz}$; see appendix~\ref{sec:local}).
}

\section{Summary and discussion}\label{sec:summary}
We have explored surface accretion driven by MHD winds as a mechanism for enhancing the dust-to-gas mass ratio in protoplanetary disks. This mechanism depletes gas at the disk surface while retaining dust concentrated near the midplane, effectively filtering the dust. This dust retention occurs when the timescale of gas removal is shorter than the timescale of dust radial transport  (equation~\eqref{eq:cond_dustretention}), thus requiring slowly drifting grains. Assuming the balance between collisional coagulation and fragmentation (equation~\eqref{eq:Stfrag}), dust retention favors poorly sticky grains with sticking threshold velocities of $\lesssim 1~\rm m~s^{-1}$, depending on turbulence strength and wind stress (equation~\eqref{eq:vstick_dustretention}). Our one-dimensional simulations have demonstrated that dust retention can indeed occur and enhance the midplane dust-to-gas mass ratio above unity  when the predicted conditions are met (figures~\ref{fig:fdg}--\ref{fig:fdg_vstick_alphawind}). Thus, dust retention by MHD-driven surface accretion may enable planetesimal formation from poorly sticky grains via the streaming and gravitational instabilities.

As described in section~\ref{sec:intro}, there are multiple lines of observational evidence suggesting that dust grains in protoplanetary disks are indeed poorly sticky and may therefore drift slowly. However, it remains to be explored whether our model can also explain other observational properties of protoplanetary disks. In a forthcoming paper, we will compare our gas and dust evolution model against disk survey observations to test our dust retention scenario in detail.

Further investigation is needed to assess the implications of our dust retention mechanism for planetesimal formation. {For $\Delta |v_{\rm g,\phi,mid}|/c_{\rm s} = 0.05$, which approximately applies to the disk models presented in this study (where $|\Delta v_{\rm g,\phi,mid}|/c_{\rm s} = 0.02$--0.05 at $r \lesssim 10$ au)}, previous simulations of the streaming instability have confirmed strong clumping of particles with ${\rm St}_{\rm mid}$ down to $10^{-3}$ \citep{Yang17,LiYoudin21}. However, our default surface accretion model with $v_{\rm stick} = 0.3~\rm m~s^{-1}$ predicts that the Stokes number of the mass-dominating grains can fall below $10^{-3}$ in regions where the dust-to-gas mass ratio exceeds unity (see the lower right panel of figure~\ref{fig:gasdust_surface}). It remains to be explored whether the streaming instability can produce strong clumping of aerodynamically well-coupled grains with ${\rm St}_{\rm mid} < 10^{-3}$ and, if so, how large the dust-to-gas ratio needs to be. We encourage future simulations of the streaming instability to explore this direction. On the other hand, we stress that efficient dust retention of ${\rm St}\sim 10^{-3}$ particles is possible if strong wind-driven accretion with $\alpha_{\rm wind} > 10^{-3}$ is present. In fact, in our surface accretion model with $v_{\rm stick} = 1~\rm m~s^{-1}$ and $\alpha_{\rm wind}  =2\times 10^{-3}$ presented in section~\ref{sec:vstick}, ${\rm St}_{\rm mid}$ exceeds $10^{-3}$ (see equation~\eqref{eq:Stfrag}), yet the midplane dust-to-gas ratio still exceeds unity. Moreover, once the dust density exceeds the gas density, aerodynamical feedback from dust to gas, which was neglected in this study, may promote dust retention in a runaway fashion \citep{Drazkowska16,IdaGuillot16,HyodoIdaGuillot21}, potentially resulting in the gravitational collapse of the dust overdensity. Our future modeling will include this feedback effect to assess the viability of planetesimal formation via the dust retention mechanism. {We also plan to combine our gas--dust evolution model with linear stability analysis of the streaming instability \citep[e.g.,][]{YoudinGoodman05} to semi-analytically predict the growth rate of the streaming instability in surface-accreting disks.}

Finally, we emphasize that the primary objective of this paper was to introduce the basic concept of dust retention in surface-accreting disks, not to present a gas and dust evolution model that accounts for the detailed vertical structures of MHD-driven accretion flows. For this purpose, we have employed a relatively simple disk model, assuming a surface accretion flow driven by MHD wind stresses along with a vertically uniform flow driven by viscosity. However, our simple model already predicts that the gas accretion velocity at the midplane, where settled dust resides, is a critical factor for dust retention. Existing global non-ideal MHD simulations \citep[e.g.,][]{Bai16,Lesur21,Iwasaki24} suggest that whether a strong accretion flow occurs at the surface or the midplane depends on the vertical ionization profile of the disk, and on the polarity of the global poloidal magnetic field when the Hall effect is significant \citep{Bai16}. Constructing an MHD-based model to predict the vertical flow structure as a function of the ionization profile, magnetic field strength, and magnetic field polarity is beyond the scope of this work but is worth pursuing in future work.

\begin{ack}
The author thanks Takahiro Ueda, Shoji Mori, Masahiro Ikoma, Kazumasa Ohno, Ryosuke Tominaga, and Yuya Fukuhara for useful discussions, and the anonymous referee for constructive comments. 
\end{ack}

\section*{Funding}
This work was supported by JSPS KAKENHI Grant Numbers JP19K03926, JP20H00182, JP20H01948, JP23H01227, and JP23K25923.

\bibliographystyle{apj}
\bibliography{SurfaceAccretion}

\appendix 

\section{Efficiency of dust transport by a localized surface accretion low}\label{sec:local}
In this study, we have assumed that the surface accretion flow does not transport settled dust. 
Here, we validate this assumption by explicitly considering an accretion flow localized at height $z = z_{\rm s}$. 

We model the localized flow as
\beq
v_{{\rm g},r}(z) = \frac{\brac{v_{{\rm g},r}}_{\rm g}\Sigma_{\rm g}}{\rho_{\rm g}(z_{\rm s})}
\delta(z-z_{\rm s}),
\label{eq:vgrwind_layer}
\eeq
where $\delta(z-z_{\rm s})$ is the delta function peaked at $z=z_{\rm s}$.
The prefactor $\brac{v_{{\rm g},r}}_{\rm g}\Sigma_{\rm g}/{\rho_{\rm g}(z_{\rm s})}$ guarantees that the definition $\brac{v_{{\rm g},r}}_{\rm g}
\equiv (1/\Sigma_{\rm g})\int v_{{\rm g},r}(z) \rho_{\rm g}(z) dz$ (see equation~\eqref{eq:ovlX}) is satisfied.
Using equation~\eqref{eq:vgrwind_layer}, the co-accretion velocity (the first term on the right-hand side of equation~\eqref{eq:vdravr}) reduces to
equation~\eqref{eq:coacc_surface}, with $C_{\rm surfac}$ given by equation~\eqref{eq:Csurface_def}.

To proceed further, we assume a balance between dust settling and vertically uniform turbulent diffusion. We also assume that the grains obey Epstein's drag law, $t_{\rm stop}(z) \propto 1/\rho_{\rm g}(z)$. Using $\rho_{\rm g} \propto \exp[-z^2/(2H_{\rm g}^2)]$, the equilibrium vertical dust distribution approximates to \citep{Takeuchi02,Fukuhara21}
\begin{align}
\rho_{\rm d}(z) &\approx \frac{\Sigma_{\rm d}}{\sqrt{2\pi}H_{\rm d}} 
\exp\left[
-\frac{z^2}{2H_{\rm g}^2}-\frac{{\rm St}_{\rm mid}}{\alpha_{Dz}}\left(\exp\frac{z^2}{2H_{\rm g}^2}-1\right)\right]
\notag \\
&=
\frac{\Sigma_{\rm d}}{\sqrt{2\pi}H_{\rm d}} 
\exp\left[
-\frac{z^2}{2H_{\rm g}^2}-\frac{{\rm St}(z)-{\rm St}_{\rm mid}}{\alpha_{Dz}}\right],
\label{eq:rhod_Takeuchi}
\end{align}
with $H_{\rm d}$ given by equation~\eqref{eq:Hd}. Here, ${\rm St}(z) \equiv {\rm St}_{\rm mid}\exp[z^2/(2H_{\rm g}^2)]$ is the Stokes number at height $z$. 
At $z \lesssim H_{\rm g}$, equation~\eqref{eq:rhod_Takeuchi} approximates to equation \eqref{eq:rhod_Gaussian} used in the main text.

\begin{figure}[t]
\begin{center}
\includegraphics[width=\hsize, bb=0 0 288 216]{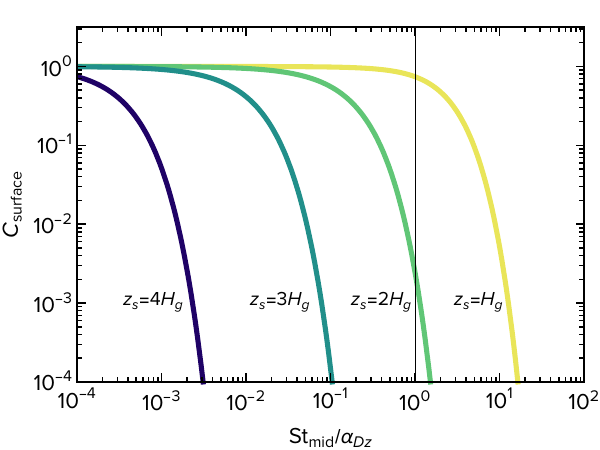}
\end{center}
\caption{Dimensionless efficiency $C_{\rm surface}$ of dust transport by a localized surface gas flow (equation~\eqref{eq:Csurface}) as a function of ${\rm St}_{\rm mid}/\alpha_{Dz}$ for different values of the flow height $z_{\rm s}$ above the midplane.
}
\label{fig:Csurface}
\end{figure}
The exponential factor $\exp[-{\rm St}(z)/\alpha_{Dz}]$ in equation~\eqref{eq:rhod_Takeuchi} indicates that  grains are heavily depleted at heights $z$ where ${\rm St}(z) \gg \alpha_{Dz}$.
Therefore, we expect that $C_{\rm surface}$ should vanish if ${\rm St}(z_{\rm s}) \gg \alpha_{Dz}$.
In fact, inserting equations~\eqref{eq:rhog_Gaussian} and \eqref{eq:rhod_Takeuchi} into equation \eqref{eq:Csurface_def} yields
\beq
C_{\rm surface} \approx \frac{H_{\rm g}}{H_{\rm d}}
\exp\left[-\frac{{\rm St}(z_{\rm s})-{\rm St}_{\rm mid}}{\alpha_{Dz}}\right],
\label{eq:Csurface}
\eeq
which confirms that $C_{\rm surface}$ decreases exponentially with increasing ${\rm St}(z_{\rm s})/\alpha_{Dz}$. 
Figure~\ref{fig:Csurface} plots $C_{\rm surface}$ as a function of ${\rm St}_{\rm mid}/\alpha_{Dz}$ for different values of $z_{\rm s}$. 
For $z_{\rm s} = 2H_{\rm g}$ and $4H_{\rm g}$, $C_{\rm surface}$ falls below 0.1 at ${\rm St}_{\rm mid}/\alpha_{Dz} \gtrsim 0.3$ and $10^{-3}$, respectively.
Therefore, MHD wind-driven accretion flows localized at $z \sim 2$--$4H_{\rm g}$ (see the references in section~\ref{sec:idea}) would not efficiently transport grains with ${\rm St}_{\rm mid} > \alpha_{Dz}$.

\section{Vertical integration of the radial diffusion term for constant $D_{{\rm d},r}$} \label{sec:diff}

Here, we perform the vertical integration of the radial diffusion term in equation~\eqref{eq:rhod} for the special case of vertically constant radial diffusion coefficient $D_{{\rm d},r}$. As shown in equation~\eqref{eq:Sigmad0}, the vertically integrated diffusion term can be written as 
$-\Sigma_{\rm g}D_{{\rm d},r}\brac{(\rho_{\rm d}/\rho_{\rm g})'}_{\rm g}$, where the prime $'$ denotes a radial partial derivative. To evaluate $\brac{(\rho_{\rm d}/\rho_{\rm g})'}_{\rm g}$ analytically, we assume the Gaussian vertical distributions of gas and dust already used in the main text (equations~\eqref{eq:rhog_Gaussian} and \eqref{eq:rhod_Gaussian}). 

We begin by explicitly writing down the vertically integrated diffusion term, 
\begin{align}
& -D_{{\rm d},r}\Sigma_{\rm g}\left\langle \pfrac{\rho_{\rm d}}{\rho_{\rm g}}' \right\rangle_{\rm g} 
\notag \\
& = -\frac{D_{{\rm d},r}\Sigma_{\rm g}}{\sqrt{2\pi}H_{\rm g}}
\int e^{-z^2/(2H_{\rm g}^2)}
\left(\frac{\Sigma_{\rm d}}{\Sigma_{\rm g}}\frac{H_{\rm g}}{H_{\rm d}}e^{-z^2/(2H_{\rm dg}^2)}\right)' 
dz,
\end{align}
where $H_{\rm dg} \equiv (H_{\rm d}^{-2}-H_{\rm g}^{-2})^{-1/2}$ is the scale height for the dust-to-gas ratio $\rho_{\rm d}(z)/\rho_{\rm g}(z)$.
We expand the integrand as 
\begin{align}
& e^{-z^2/(2H_{\rm g}^2)}
\left(\frac{\Sigma_{\rm d}}{\Sigma_{\rm g}}\frac{H_{\rm g}}{H_{\rm d}}e^{-z^2/(2H_{\rm dg}^2)}\right)'
\notag \\
&= e^{-z^2/(2H_{\rm d}^2)}\left[\frac{H_{\rm g}}{H_{\rm d}}
\left(\frac{\Sigma_{\rm d}}{\Sigma_{\rm g}}\right)' 
+ \frac{\Sigma_{\rm d}}{\Sigma_{\rm g}}
\left(\frac{H_{\rm g}}{H_{\rm d}}\right)'
+ \frac{z^2}{H_{\rm dg}^3} \frac{\Sigma_{\rm d}}{\Sigma_{\rm g}}\frac{H_{\rm g}}{H_{\rm d}}H_{\rm dg}'
\label{eq:dfdgdr}
\right].
\end{align}
Performing vertical integration, we have 
\begin{align}
& \frac{1}{\sqrt{2\pi}}\int e^{-z^2/(2H_{\rm g}^2)}
\left(\frac{\Sigma_{\rm d}}{\Sigma_{\rm g}}\frac{H_{\rm g}}{H_{\rm d}}e^{-z^2/(2H_{\rm dg}^2)}\right)'dz
\notag \\
&= H_{\rm d}\left[\frac{H_{\rm g}}{H_{\rm d}}
\left(\frac{\Sigma_{\rm d}}{\Sigma_{\rm g}}\right)' 
+ \frac{\Sigma_{\rm d}}{\Sigma_{\rm g}}
\left(\frac{H_{\rm g}}{H_{\rm d}}\right)'
+ \frac{H_{\rm d}^2}{H_{\rm dg}^3} \frac{\Sigma_{\rm d}}{\Sigma_{\rm g}}\frac{H_{\rm g}}{H_{\rm d}}H_{\rm dg}'
\right]
\notag \\
&= H_{\rm g}\left[
\left(\frac{\Sigma_{\rm d}}{\Sigma_{\rm g}}\right)' 
+ \frac{\Sigma_{\rm d}}{\Sigma_{\rm g}}
\left(
\frac{H_{\rm g}'}{H_{\rm g}}
- \frac{H_{\rm d}'}{H_{\rm d}}
+ \frac{H_{\rm d}^2}{H_{\rm dg}^3} H_{\rm dg}' \right)
\right]
\notag \\
&= H_{\rm g}\left[
\left(\frac{\Sigma_{\rm d}}{\Sigma_{\rm g}}\right)' 
+ \frac{\Sigma_{\rm d}}{\Sigma_{\rm g}}
\left(
1-\pfrac{H_{\rm d}}{H_{\rm g}}^2 \right)
(\ln H_{\rm g})'
\right].
\end{align}
Hence, we obtain the final result
\begin{align}
-D_{{\rm d},r}\Sigma_{\rm g}\left\langle\pfrac{\rho_{\rm d}}{\rho_{\rm g}}' \right\rangle_{\rm g} 
=&
-D_{{\rm d},r}\Sigma_{\rm g}\left(\frac{\Sigma_{\rm d}}{\Sigma_{\rm g}}\right)'
\notag \\
& - D_{{\rm d},r}
\left(
1-\pfrac{H_{\rm d}}{H_{\rm g}}^2 \right)
(\ln H_{\rm g})'
\Sigma_{\rm d}.
\label{eq:diffflux_final}
\end{align}
On the right-hand side of equation~\eqref{eq:diffflux_final}, the first term represents the radial diffusion term for $\Sigma_{\rm d}$, which is commonly adopted in the literature. Our new discovery here is the second term. This term vanishes when $H_{\rm d} = H_{\rm g}$, but remains otherwise. If 
$\Sigma_{\rm d}/\Sigma_{\rm g}$ and $H_{\rm g}$ vary over a length scale of $\sim r$, and if $H_{\rm d} \ll H_{\rm g}$, then the second term is comparable in the magnitude to the first term. Importantly, the second term is proportional to $\Sigma_{\rm d}$, indicating that it behaves as an {\it advection} term. The corresponding advection velocity is $-D_{{\rm d},r}( 1-(H_{\rm d}/H_{\rm g})^2 ) (\ln H_{\rm g})'$. Typically, $(\ln H_{\rm g})'$ is positive, and hence this advection velocity is negative, transporting dust inward.

\begin{figure}[t]
\begin{center}
\includegraphics[width=\hsize, bb=0 0 260 167]{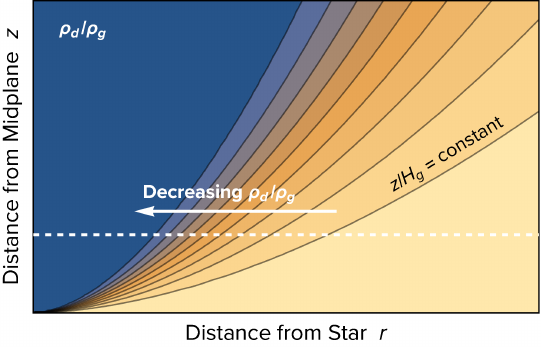}
\end{center}
\caption{Schematic illustration showing  how a vertically stratified dust-to-gas density ratio $\rho_{\rm d}/\rho_{\rm g}$ produces a radial dust diffusion flux even when the surface density ratio $\Sigma_{\rm d}/\Sigma_{\rm g}$ is radially constant. Shown are isolines of $\rho_{\rm d}/\rho_{\rm g}$ in the $r$--$z$ plane for a disk with radially constant $\Sigma_{\rm d}/\Sigma_{\rm g}$ and $H_{\rm d}/H_{\rm g} (<1)$ but radially increasing $H_{\rm g}$. Above the midplane (dashed line), $\rho_{\rm d}/\rho_{\rm g}$ decreases with decreasing $r$, yielding a dust diffusion flux toward the central star.
}
\label{fig:Dadvection}
\end{figure}
This apparent advection term arises because, even when $\Sigma_{\rm d}/\Sigma_{\rm g}$ is radially constant, $\rho_{\rm d}/\rho_{\rm g}$ still has a positive radial gradient if $H_{\rm g}$ increases with $r$ and if $\rho_{\rm d}/\rho_{\rm g}$ is vertically stratified ($H_{\rm d} < H_{\rm g}$). This can be understood by drawing isolines of $\rho_{\rm d}/\rho_{\rm g}$ in the $r$--$z$ plane for the special case of radially constant $\Sigma_{\rm d}/\Sigma_{\rm g}$ and $H_{\rm d}/H_{\rm g}$ but radially increasing $H_{\rm g}$ (figure~\ref{fig:Dadvection}). In this case, the isolines align with lines of constant $z/H_{\rm g}(r)$. At the midplane, $\rho_{\rm d}/\rho_{\rm g}$ is radially constant, yielding no radial diffusion flux. However, above the midplane, the heights of all isolines increase with $r$, indicating that $\rho_{\rm d}/\rho_{\rm g}$ has a nonzero radial gradient. Since $\rho_{\rm d}/\rho_{\rm g}$ above the midplane decreases toward the central star, the radial diffusion flux there is inward.

\end{document}